\documentclass[12pt]{iopart}
\expandafter\let\csname equation*\endcsname=\relax
\expandafter\let\csname endequation*\endcsname=\relax
\usepackage{amsmath, amssymb, color, graphicx, stmaryrd, wasysym}
\definecolor{linkcolor}{rgb}{0,0,0.6} 
\usepackage[colorlinks=true,
	pdfstartview = FitV,
	linkcolor    = linkcolor,
	citecolor    = linkcolor,
	urlcolor     = linkcolor,
	hyperindex   = true,
	hyperfigures = false]{hyperref}

\usepackage{amsmath,amssymb,graphicx,cases}
\usepackage{epsfig}
\usepackage{textcomp}
\usepackage{epstopdf}
\usepackage{float}
\usepackage{braket}
\usepackage[normalem]{ulem}
\usepackage{enumerate}
\usepackage{enumitem}
\usepackage{ae,aecompl}
\usepackage{lmodern}

\usepackage{xcolor,cancel}
\definecolor{mygreen}{rgb}{0.0,0.55,0.4}

\newcommand{\stkout}[1]{\ifmmode\text{\sout{\ensuremath{#1}}}\else\sout{#1}\fi}

\def\\{\textcolor{blue}}

\DeclareMathOperator{\arctanh}{arctanh}

\begin{document}

\title[Thermodynamically consistent flocking]{Thermodynamically consistent flocking: From discontinuous to continuous transitions}

\author{Tal Agranov}
\address{DAMTP, Centre for Mathematical Sciences, University of Cambridge, Wilberforce Road, Cambridge CB3 0WA, United Kingdom}
\ead{tal.agranov@mail.huji.ac.il}

\author{Robert L. Jack}
\address{DAMTP, Centre for Mathematical Sciences, University of Cambridge,
Wilberforce Road, Cambridge CB3 0WA, United Kingdom}
\address{Yusuf Hamied Department of Chemistry, University of Cambridge, Lensfield Road, Cambridge CB2 1EW, United Kingdom}

\author{Michael E. Cates}
\address{DAMTP, Centre for Mathematical Sciences, University of Cambridge, Wilberforce Road, Cambridge CB3 0WA, United Kingdom}

\author{\'Etienne Fodor}
\address{Department of Physics and Materials Science, University of Luxembourg, L-1511 Luxembourg, Luxembourg}

\begin{abstract}
We introduce a family of lattice-gas models of flocking, whose thermodynamically consistent dynamics admits a proper equilibrium limit at vanishing self-propulsion. These models are amenable to an exact coarse-graining which allows us to study their hydrodynamic behavior analytically. We show that the equilibrium limit here belongs to the universality class of Model C, and that it generically exhibits tricritical behavior. Self-propulsion has a non-perturbative effect on the phase diagram, yielding novel phase behaviors depending on the type of aligning interactions. For aligning interaction that increase monotonically with the density, the tricritical point diverges to infinite density reproducing the standard scenario of a discontinuous flocking transition accompanied by traveling bands. In contrast, for models where the aligning interaction is non-monotonic in density, the system can exhibit either (the nonequilibrium counterpart of) an azeotropic point, associated with a continuous flocking transition, or a state with counterpropagating bands.
\end{abstract}

\maketitle


\section{Introduction}

Flocking is a prominent phenomenon in nonequilibrium statistical mechanics~\cite{toner_hydrodynamics_2005, Chate2020}, comprising the collective motion of a large group of aligning self-propelled agents. It appears in systems ranging from bird flocks~\cite{Giardina2014} to human crowds~\cite{Bartolo2019} and synthetic self-propelled colloids~\cite{Bartolo2013}. Various theoretical models lead to a qualitatively similar flocking behavior, which suggests possible universality. A pioneering theoretical account of flocking is due to Vicsek {\em et al.}~\cite{vicsek_novel_1995}, who proposed an agent-based model almost three decades ago, followed by Toner and Tu~\cite{toner_long-range_1995} who offered a hydrodynamic description. Since then, these models and their variants have played a leading role in the study of active matter~\cite{marchetti_hydrodynamics_2013}.

In the Vicsek model, locally aligning spins move with fixed speed and individually fluctuating orientation, which drives the dynamics out of equilibrium. Strikingly, under local alignment, this model undergoes an ordering transition even in two dimensions, which would be precluded in equilibrium by the Mermin-Wagner theorem~\cite{Mermin1966, Halperin2019}. When increasing either density or alignment, the system transitions from a homogeneous disordered gas to a state of homogeneous collective motion with long-ranged order. Originally thought of as a continuous transition~\cite{vicsek_novel_1995}, it took almost another decade to establish flocking as a discontinuous transition~\cite{gregoire_onset_2004}: in between the two homogeneous phases, there is a region in parameter space where dense ordered bands propagate against a dilute disordered background. Unlike equilibrium liquid-gas transitions, the critical point is at infinite density. This feature has been attributed to the fact that the dense and dilute phases have different symmetries, so that they cannot be connected via a continuous transformation~\cite{solon_flocking_2015}.

A recent lattice-based model, known as the active Ising model (AIM), has led to a series of analytical predictions for the phase diagram of the flocking transition~\cite{Solon2013, solon_pattern_2015}. In this simple extension of the Ising model, motile spins with discrete symmetry can only self-propel along a single axis. Importantly, this model is amenable to coarse-graining, yielding an exact hydrodynamic description~\cite{solon_flocking_2015, kourbane-houssene_exact_2018-1, scandolo_active_2023}. Although the phenomenology of AIM is clearly distinct from any equilibrium analogue, it is tempting to try and rationalize how AIM departs from an equilibrium reference model, with proper reversible dynamics, as was done for repulsive active particles~\cite{Fodor2016, Fodor2022}. Interestingly, AIM does not reduce to equilibrium at vanishing self-propulsion, as one might expect. This inconsistency reveals that AIM is actually not compatible with the theory of stochastic thermodynamics~\cite{seifert_stochastic_2012}, which assumes that the system is in contact with equilibrium reservoirs and driven by external forces (so that it naturally relaxes to equilibrium in the absence of drive). The same holds for other flocking models with continuous spin symmetry~\cite{vicsek_novel_1995}, or with topological interactions that involve nonreciprocal alignment~\cite{ballerini_interaction_2008, ginelli_relevance_2010}, which also lead to irreversibility even at vanishing self-propulsion.

It is then natural to examine flocking models where the microscopic dynamics is amended to entail an equilibrium limit in the absence of self-propulsion, which we refer to as {\em thermodynamic consistency}. Several questions remain open: Would novel qualitative behavior emerge from thermodynamic consistency? To which universality class does the equilibrium limit of such flocking models belong? Inspired by AIM, we explore these issues in the context of flocking systems with a discrete symmetry. To make analytical progress, we choose lattice models which, like AIM, admit an analytically exact and tractable hydrodynamic description.

This work is part of a broader current effort in recasting various active systems in a thermodynamically consistent framework such as in phase field crystal models \cite{Vrugt2022}, active fluids \cite{fischer_aggregation_2019,bebon_thermodynamics_2024} , and   reaction-diffusion systems \cite{aslyamov_nonideal_2023}. Although our models are permanently out of  thermodynamic equilibrium, we find that their behavior can be usefully analyzed in terms of the underlying free energy that controls their equilibrium limit at vanishing self-propulsion. Indeed, the free energy plays an informative role even outside the regime of linear response to the nonequilibrium drive. As noted previously~\cite{martin_fluctuation-induced_2021,Solon2013,solon_flocking_2015}, the standard scenario of a discontinuous flocking transition stems from a density modulated coupling of the alignment interaction. In our models, we reveal a strikingly different phenomenology between the cases of monotonic and non-monotonic density modulations. For models where ordering increases monotonically with density, the equilibrium limit exhibits a single tricritical point~\cite{scandolo_active_2023}, and introducing activity is a non-perturbative effect: the tricritical point diverges to infinite density, and the flocking transition is discontinuous all across phase space. In contrast, for models where ordering is non-monotonic with density, the equilibrium limit exhibits a pair of tricritical points. At finite self-propulsion, such points can either collide into a single azeotropic point, yielding a continuous flocking transition, or instead a new collective state emerges with counterpropagating bands.

This paper is organized as follows. In Sec.~\ref{sec:def}, we set up the general definition of our thermodynamically consistent flocking models. We propose a microscopic dynamics with a generic type of aligning interactions, and derive the corresponding hydrodynamic description through systematic coarse-graining. Then, we examine in Secs.~\ref{sec:mono} and~\ref{sec.non.mon} the phase diagrams for monotonic and non-monotonic dependence of the aligning strength on density, respectively. Overall, our results illustrate how various choices of density-dependent alignment affect the flocking transition, yielding unprecedented phenomenology in the flocking context. Yet, some of this phenomenology echoes that of equilibrium systems, so that our thermodynamically consistent description allows one to anticipate and classify the topologies of nonequilibrium phase diagrams in terms of their equilibrium counterparts.


\section{Thermodynamic consistency: From microscopics to hydrodynamics}\label{sec:def}

In this Section, we introduce a class of lattice dynamics for active spins that obey thermodynamic consistency. We take locally aligning interactions with mesoscopic range to capture a transition to collective motion. Such a form of interactions is amenable to exact coarse-graining, which allows us to derive the corresponding hydrodynamic equations for some relevant fields. Importantly, our derivation is given in a functional form with an arbitrary dependence of the alignment strength on density.

\subsection{Microscopic dynamics of aligning active spins}

\begin{figure}
	\centering
	\includegraphics[width=.8\columnwidth]{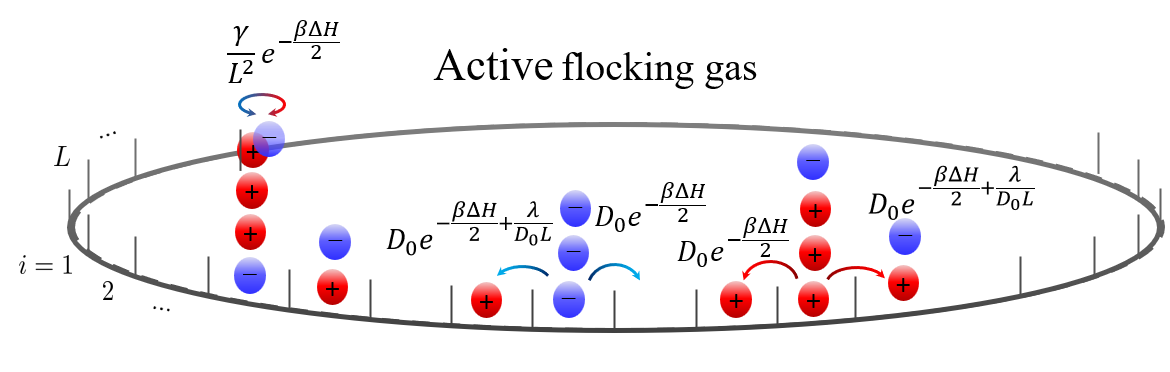}
	\caption{The active flocking model with thermodynamic consistency. Each particle can be in either one of two states ($+$ and $-$) which determine the direction of biased diffusion (respectively to the right and left). The aligning Hamiltonian $H$ constrains both the change of particle states and their diffusive hops.}
	\label{schem}
\end{figure}

Flocks with discrete symmetry, like in AIM~\cite{solon_flocking_2015}, are defined by two species of particles ($+$ and $-$) which hop with a bias to the right or left (respectively) on a periodic lattice of $L\gg 1$ sites, see Fig.~\ref{schem}. We consider for simplicity a one dimensional lattice, yet our fluctuating hydrodynamics is trivially extended to higher dimensions. The dynamics does not exclude multiple occupancy, so that each site $i$ can have any number of particles. The total number of particles in the system is $N=\rho_0 L$, where $\rho_0$ is the mean density.

The dynamical update rules consist of diffusive hops biased by self-propulsion, and spin flips $+\leftrightarrow-$. Differently from previous models~\cite{solon_flocking_2015, scandolo_active_2023}, at zero self-propulsion we make both these rules derive from the same Hamiltonian $H$ via a detailed balance constraint~\cite{Lebowitz1999, seifert_stochastic_2012}. We leave $H$ unspecified at this stage. Self-propulsion is then added as a weak bias of hops in the direction of the particle's spin, yielding the following update rules (see Fig.~\ref{schem}):
\begin{enumerate}
	\item Site hoping: Any particle jumps to a neighboring site with rate $D_0e^{-\frac{\beta\Delta H}{2}+\frac{\lambda}{D_0L}}$ if the jump is in the direction of its spin, and $D_0e^{-\frac{\beta\Delta H}{2}}$ otherwise. Here, $\Delta H$ is the energy difference between the configurations of the system before and after the jump.
	\item Tumbling: A $+$ ($-$) particle converts into a $-$ ($+$) particle with rate $(\gamma/L^2)e^{-\frac{\beta\Delta H}{2}}$.
\end{enumerate}
Here, $D_0$ is the bare diffusion constant, $\lambda$ is the self-propulsion strength, and $\gamma$ sets the tumble rate. The scaling of the rates with $L$ ensures that in the hydrodynamic limit ($L\to \infty$ at fixed $\rho_0$) all processes occur on diffusive time scales~\cite{kourbane-houssene_exact_2018-1, bodineau_current_2010-1}. Indeed, the time it takes for particles to traverse a macroscopic system of size $L$, either via diffusive motion or on account of self-propulsion, scales as $L^2$ which is also the time scale for tumbling events. This $L$-dependent scaling of the rates  is standard for other lattice-based diffusive models~\cite{kourbane-houssene_exact_2018-1, bodineau_current_2010-1}. The main difference with AIM~\cite{Solon2013, solon_flocking_2015, martin_fluctuation-induced_2021} is that our site hopping now accounts for the energy difference $\Delta H$. This feature renders our models reversible at vanishing self-propulsion, thus setting a proper equilibrium limit, as a consequence of thermodynamic consistency~\cite{Lebowitz1999, seifert_stochastic_2012}.

We now restrict $H$ such that the interaction range spans a mesoscopic scale $\Delta x = L^\delta$, with $0<\delta<1$. This scale is sub-extensive in system size, yet contains a large number of particles $1\ll\Delta x\ll L$, and is the scale appearing in the coarse-graining procedure detailed below. Within this choice of scaling each particle interacts with a vanishing fraction of the system of the order of $\sim\Delta x/L\ll1$, yet it still enables a phase transition in one dimension even in the equilibrium limit. Indeed, in defiance of the usual Landau argument (which states that having only a finite energetic cost of domain walls in one dimension destroys long range order \cite{landau_impossibility_1980}) let's consider the free energy cost of the interface between phases. The configurational  entropy of a single domain wall scale as $\log L$ (as there are $L$ sites to center it upon them). Now since each particle across the domain wall interacts with about $\sim\Delta x$ other particles, then the energtic cost of a domain wall would scale at least as $\Delta x\sim L^{\delta}$, which is much larger then the entropic $\log L$ contribution (compare this with a local interaction range where the energetic cost is $\mathcal O\left(1\right)$). Thus, the large $\mathcal O\left(L^{\delta}\right)$ free energy cost will make domain walls thermodynamically unfavorable and keep the ordered state stable.

Note that it was recently found \cite{benvegnen_flocking_2022} that the flocking transition does not survive in the one-dimensional AIM. Instead it is replaced by either direction-reversing ordered aggregates or static asters composed of opposing clusters of left and right movers that block each other. These arise by virtue of the purely local interaction range of the AIM and do not appear in our model as a result of our choice of a mesoscopic interaction range.

In practice, interactions are given in terms of some local average occupancies for the density and `magnetization' variables, respectively $\eta_i^{\rho}=\eta_i^++\eta_i^- $ and $\eta_i^{m}=\eta_i^+-\eta_i^- $, where $\eta_i^{+}$ and $\eta_i^-$ are the number of $+$ and $-$ particles at site $i$. The local averages around site $i$ are given by
\begin{equation}\label{coarse}
	\hat{\rho}_i=\frac{\sum_{|i-j|<\Delta x}\eta_j^{\rho}}{2\Delta x} ,
	\quad
	\hat{m}_i=\frac{\sum_{|i-j|<\Delta x}\eta_j^{m}}{2\Delta x} .
\end{equation}
We then choose the aligning Hamiltonian to take the form
\begin{equation}\label{h}
	H = -\sum_{i=1}^L\frac{\hat{m}_i^2}{2\hat{\rho}_i}f(\hat{\rho}_i) .
\end{equation}
For $f(\hat{\rho}_i)=1$, this is the sum of $L$ fully connected Ising Hamiltonians within some restricted range $\Delta x$, where the coupling is scaled with the total number of interacting spins, see also \cite{solon_flocking_2015}. Indeed, denote by $s_i^k\in\left\{+1,-1\right\}$ with $k=1,\dots,\eta_i^{\rho}$ the spin variable of the $k$'th particle on site $i$. Then the Ising Hamiltonian with all-to-all interactions within the range $\Delta x$ of site $i$ reads
	\begin{equation}\label{local}
	-\frac{1}{2}\frac{\sum_{|i-j|<\Delta x}\sum_{|i-k|<\Delta x}\sum_{l=1}^{\eta_j}\sum_{m=1}^{\eta_k}s_j^ls_k^m}{2\Delta x\hat{\rho}_i}=-2\Delta x\frac{\hat{m}_i^2}{2\hat{\rho}_i}
	\end{equation}
The factor $1/2$ avoids double counting. The normalization $2\Delta x\hat{\rho}_i$ ensures the interactions remain extensive with the total number of interacting spins, as is common for the fully-connected Ising Hamiltonian. Then summing the right hand side \eqref{local} over all disjoint compartments of size $2\Delta x$ is equivalent to the sum \eqref{h} with $f\left(\hat{\rho}_i\right)=1$ which runs over all lattice points  (up to sub-leading corrections in $L$). Then the density-dependent modulation $f(\hat \rho_i)$, which plays a key role in the following, tunes the alignment strength. For AIM, a similar modulation appears at a coarse-grained level, as an effective renormalization due to the hydrodynamic fluctuations~\cite{martin_fluctuation-induced_2021}. In our model, since the microscopic dynamics already contains a mesoscale interaction range, we choose to directly introduce such a modulation at this level. Various forms of $f$ can be motivated on account of different types of close neighbor interactions \cite{gorbonos_pair_2020,bastien_model_2020,king_non-local_2021,wirth_is_2023}.

In the following, we consider the hydrodynamic limit of this model.  As a preliminary step, note that for a single isolated particle, there are two important length scales: $\xi_p=\lambda L/\gamma$ is the average displacement of particles between tumbles due to self-propulsion, and $\xi_d=L\sqrt{D_0/\gamma}$ is the typical diffusive displacement, over the same time.  These distances are measured in units of the lattice spacing and they are both $\mathcal{O}(L)$, because of the $L$-dependent hopping rates of the model.


\subsection{Hydrodynamic equations for density and magnetization}\label{hyd}

We follow a standard coarse-graining procedure, inspired by similar lattice models with diffusive scaling~\cite{jona-lasinio_large_1993}, which has already been deployed in some active lattice gas models~\cite{agranov_exact_2021, agranov_macroscopic_2023}. We start with a spatial coarse-graining over the diffusive displacement $\xi_d$. I.e., we  define the macroscopic spatial coordinate $x=i/\xi_d\in[0,\ell_s]$ where $\ell_s=L/\xi_d=\sqrt{\gamma/D_0}$ is the system size in the hydrodynamic representation (measured in units of $\xi_d$, as in~\cite{kourbane-houssene_exact_2018-1}). The reason for this choice is that we will later analyse phase-separated profiles: their interfacial widths are of order 1 in this hydrodynamic representation and it will be convenient to take $\ell_s\gg1$ so that the system is large compared to this interfacial width.  For consistency with these choices we make a diffusive scaling of time, setting $t=\gamma\hat{t}/L^2$ with $\hat{t}$ the time variable of the microscopic system.

The local density in the vicinity of point $x$ is defined by the mesoscopic coarse-graining~\eqref{coarse} as:
\begin{equation} \label{den}
	\rho(x,t)=\hat{\rho}_{\frac{xL}{\ell_s}}\left(t\right) ,
	\quad
	m(x,t)=\hat{m}_{\frac{xL}{\ell_s}}\left(t\right) .
\end{equation}
Since $\rho$ is conserved and $m$ is not, the dynamics of these fields must take the general form
\begin{equation}\label{d1}
	\begin{bmatrix}
	\partial_t{\rho}\\
	\partial_tm
	\end{bmatrix}=-\partial_x\begin{bmatrix}
	J_{\rho}\\
	J_m
	\end{bmatrix}
	-\begin{bmatrix}
	0\\
	2K
	\end{bmatrix}
\end{equation} 
where $J_{\rho}, J_m$ are the conservative fluxes and $K$ is the net rate of change of the $\rho_-$ density due to spin flipping. As detailed in~\ref{ap.hydro}, to leading order in system size we find the following expressions
\begin{equation}\label{d2}
	\begin{bmatrix}
	J_{\rho}\\
	J_m
	\end{bmatrix}=-\frac{1}{2}\mathbb C(\rho,m)\partial_x\begin{bmatrix}
	\frac{\delta F}{\delta\rho}\\
	\frac{\delta F}{\delta m}
	\end{bmatrix}+\text{Pe}\begin{bmatrix}
	m\\
	\rho
	\end{bmatrix},
	\quad
	K=M(\rho,m) \sinh\left(\frac{\delta F}{\delta m}\right),
	\quad
	\text{Pe}=\frac{\lambda}{\sqrt{D_0\gamma}} .
\end{equation} 
The P\'eclet number $\text{Pe}=\xi_p/\xi_d$ is the standard measure of activity in similar active systems with diffusive scaling~\cite{agranov_entropy_2022}. Here, $F[\rho,m]=\frac{L}{\ell_s}\int_0^{\ell_s}dx\mathcal F[\rho(x,t),m(x,t)]$ is the free-energy functional of the equilibrium ($\lambda=0$) system with the corresponding free energy density given by the function 
\begin{equation}\label{free}
	\mathcal F \left(\rho,m\right)= \frac{\rho}{2}\ln\frac{\rho^2-m^2}{4}+\frac{m}{2}\ln\frac{\rho+m}{\rho-m}-\rho-\beta \frac{m^2}{2\rho}f\left(\rho\right),
\end{equation}	
while $\mathbb C$ and $M$ are mobility coefficients given by 
\begin{eqnarray}\label{mob}
	\mathbb C=2\begin{bmatrix} \rho&m
	\\
	m&\rho\end{bmatrix},
	\quad
	M=\sqrt{\rho^2-m^2}.
\end{eqnarray}
The expressions~(\ref{d2}-\ref{free}) are exact as long as $\text{Pe}\neq0$. Yet in order to describe correctly the equilibrium ($\text{Pe}=0$) hydrodynamics for non convex $\mathcal F$, one has to retain surface tension terms that are sub-leading $\mathcal O\left(L^{-2}|\nabla\rho|^2,L^{-2}|\nabla m|^2\right)$. Still, as is usual in the thermodynamics of phase separation, the bulk term $\mathcal F$ suffices for determining the binodal (coexisting) densities \footnote{The profiles presented in the following correspond to finite $\text{Pe}\neq0$ and were computed without the need for the sub-leading interfacial tension terms.}.

The dynamics~(\ref{d1}-\ref{d2}) has a thermodynamically consistent structure. Indeed, for $\text{Pe}=0$, all terms derive from the free energy $\mathcal F$, which then serves as a Lyapunov function, as discussed in~\ref{ap.hydro}. As shown in what follows, this equilibrium limit belongs to the universality class of Model C~\cite{scandolo_active_2023, Bray2002}. Note that, as written, the dynamics~(\ref{d1}-\ref{d2}) neglects some sub-leading noise terms of order $L^{-1/2}$; these can be readily deduced from the detailed-balance condition (and are included by definition in Model C).

The general structure of~(\ref{d1}-\ref{d2}) should hold rather broadly for all thermodynamically consistent flocking models with discrete symmetry and under diffusive scaling, including models with purely local interaction range (albeit with a renormalized free energy depending on dimensionality). The conservative fluxes in~\eqref{d2} are linear in the free-energy derivatives, as is common when expanding close to equilibrium. That these expressions hold also far away from equilibrium is a consequence of the $L$-dependent scaling of the microscopic rates, and the diffusive scaling of space and time; any higher order spatial gradients will be sub-leading in system size $L^{-1}$. This reasoning does not apply for the flipping term $K$, since it does not involve spatial variations of the fields. Indeed, this term goes beyond linear order in the free energy derivatives even at the hydrodynamic level~\cite{kaiser_canonical_2018-1}. Activity ($\text{Pe}>0$) leads to the extra terms $\text{Pe}\, m$ and $\text{Pe}\,\rho$ in the conservative fluxes~\eqref{d2} but otherwise does not interfere with the remaining bare equilibrium terms. Again, this is a consequence of the diffusive scaling, where self-propulsion enters as a weak bias in the microscopic dynamics.

Our thermodynamically consistent dynamics in~(\ref{d1}-\ref{d2}) is easily related to the non-consistent one of AIM~\cite{Solon2013, solon_flocking_2015, martin_fluctuation-induced_2021}. Our flipping term $K$ is the same as in the hydrodynamic equations of AIM, under appropriate matching of $f$. The conservative fluxes $J_\rho$ and $J_m$ are different from those of AIM, since they stem from a different choice of hopping rules. Nevertheless, one can reproduce the corresponding terms in AIM by sending the temperature that enters these terms to infinity (namely, taking $\beta=0$ in the free energy that enters the conservative fluxes $J_\rho$ and $J_m$), as discussed in~\ref{ap.hydro}.
Indeed, denote by $\mathcal{F}_{0}$ the free energy density evaluated at $\beta=0$ and write $F_0$ for the corresponding free energy functional.  Then the hydrodynamics of the AIM are also given by (\ref{d1},\ref{d2}), except that one replaces $F\to F_0$ in the expressions for $J$
	\begin{equation}\label{d3}
	\begin{bmatrix}
	J_{\rho}\\
	J_m
	\end{bmatrix}_{\text{AIM}}=-D_0\partial_x\begin{bmatrix}
	\rho\\
	m
	\end{bmatrix}+\text{Pe}\begin{bmatrix}
	m\\
	\rho
	\end{bmatrix}=-\frac{1}{2}\mathbb C(\rho,m)\partial_x\begin{bmatrix}
	\frac{\delta F_0}{\delta\rho}\\
	\frac{\delta F_0}{\delta m}
	\end{bmatrix}+\text{Pe}\begin{bmatrix}
	m\\
	\rho
	\end{bmatrix},
	\end{equation} while retaining the full $F$ in $K$. To this extent, the source of residual irreversibly in the zero self-propulsion limit of AIM can be regarded as a {\em mismatch between two temperatures} controlling respectively the conservative fluxes $(J_\rho, J_m)$ and the flipping term $K$.


\section{Phase diagram for monotonic alignment strength}\label{sec:mono}

Various forms of the density-dependent alignment strength $f\left(\rho\right)$, entering in the microscopic alignment~\eqref{h} and in the hydrodynamic free energy~\eqref{free}, can be motivated to reflect different types of close neighbor interactions~\cite{gorbonos_pair_2020, bastien_model_2020, king_non-local_2021,wirth_is_2023}. We first explore the phase behavior for a monotonic $f(\rho)$ by considering the form appearing previously in AIM~\cite{Solon2013, solon_flocking_2015, martin_fluctuation-induced_2021}:
\begin{equation}\label{f0}
	f\left(\rho\right)=1-\frac{r}{\rho} .
\end{equation}
This choice is convenient mathematically, although on physical grounds one might instead favor choices that are positive everywhere, and do not diverge at $\rho\to0$. Yet this does not change the phenomenology reported below. Other monotonic forms have been recently proposed in extensions of AIM yielding a similar phase behavior~\cite{scandolo_active_2023}. In some of these models, the alignment term is not purely quadratic in the magnetization as assumed here (see~\eqref{h}). Yet, its expansion close to criticality reduces to this form, and the topology of the phase diagram actually follows from the behavior close to criticality.

\begin{figure}[b]
	\centering
	\includegraphics[width=.32\columnwidth]{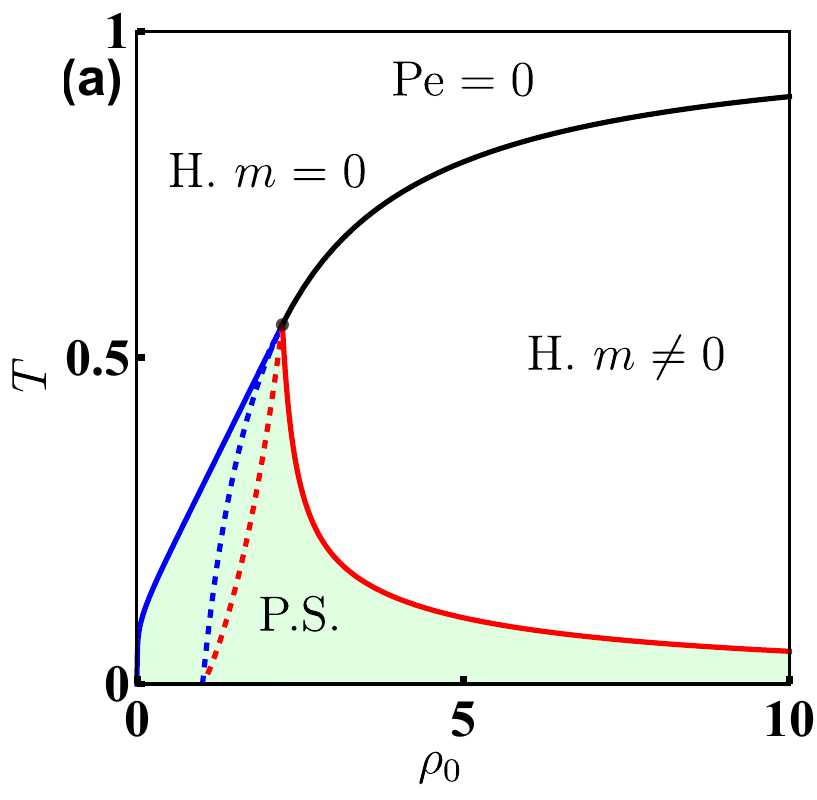}
	\includegraphics[width=.32\columnwidth]{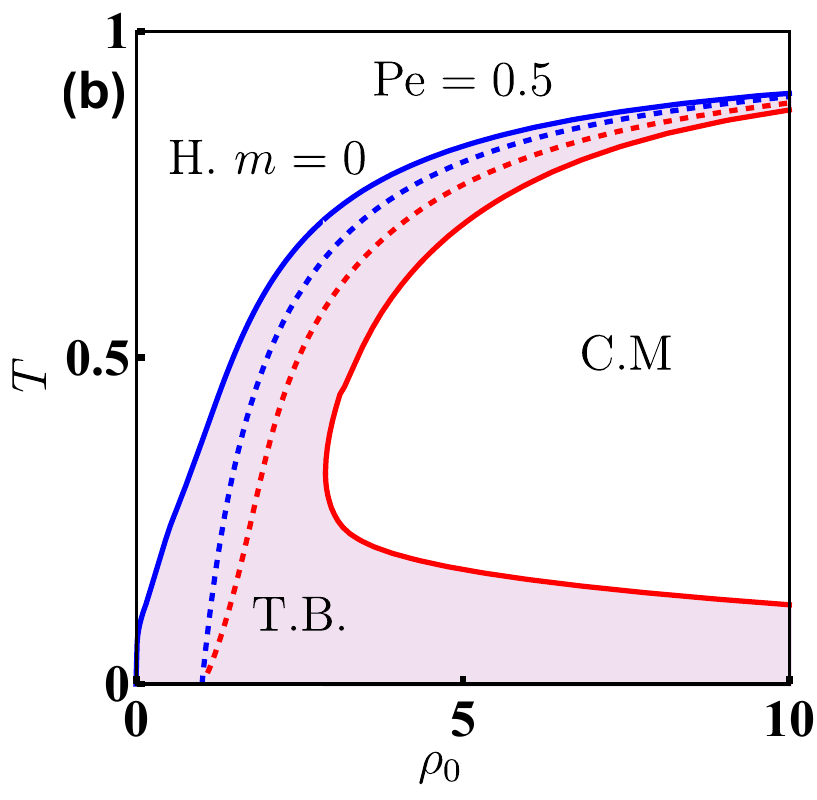}
	\includegraphics[width=.305\columnwidth]{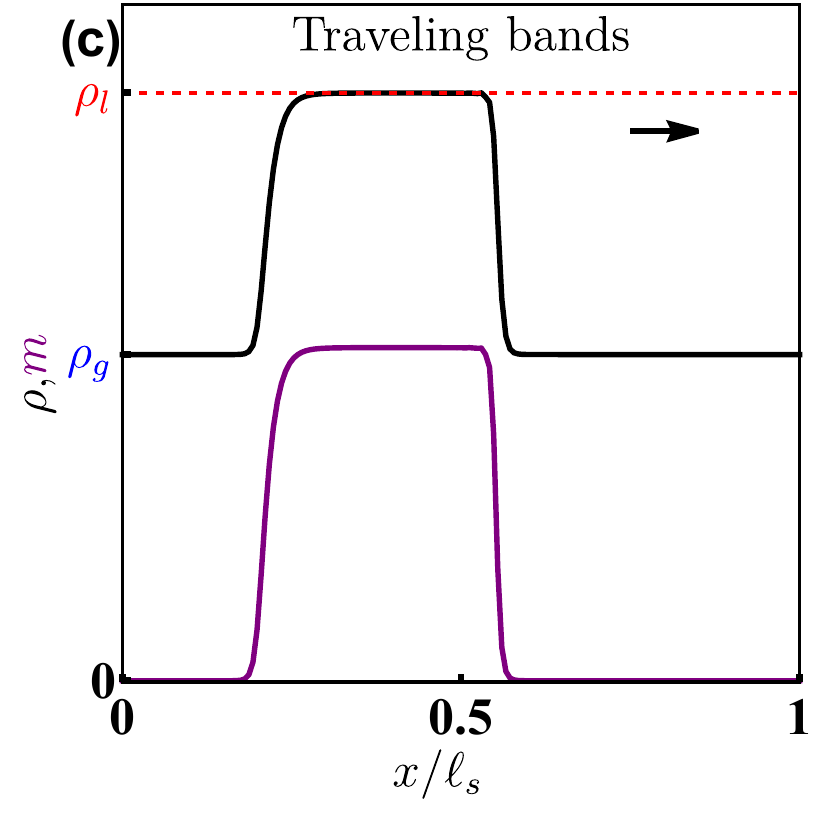}
	\caption{Phase diagrams for (a)~the equilibrium case ($\text{Pe}=0$) and (b)~the nonequilibrium regime ($\text{Pe}=0.5$), with monotonic $f$ given by~\eqref{f0} with $r=1$. The colored dashed and solid curves are spinodals and binodals, respectively. The black solid curve is a critical line. The black dot marks a tricritical point. (c)~Traveling band (T.B.) profiles for $\text{Pe}=0.5$, $\rho_0\simeq3.62$, and $T\simeq0.71$.}
	\label{fig.phase1}
\end{figure}

The resulting phase diagram, given in terms of $T=1/\beta$ and $\rho_0$, is composed of three phases shown in Fig.~\ref{fig.phase1}: (i)~a homogeneous disordered phase (H; $m=0$), (ii)~a homogeneous ordered phase (H; $m\neq0$), which becomes a collective motion (C.M.) for non-vanishing self-propulsion, and (iii)~a phase-separated state (P.S.) between different discrete symmetries ($m=0$ and $m\neq0$), which becomes a traveling  band (T.B.) for non-vanishing self-propulsion. In this Section, we study the topology of this phase diagram and derive its phase boundaries. We address first the equilibrium limit ($\text{Pe}=0$), and then the nonequilibrium regime ($\text{Pe}>0$).


\begin{figure}[b]
	\centering
	\includegraphics[width=.32\columnwidth]{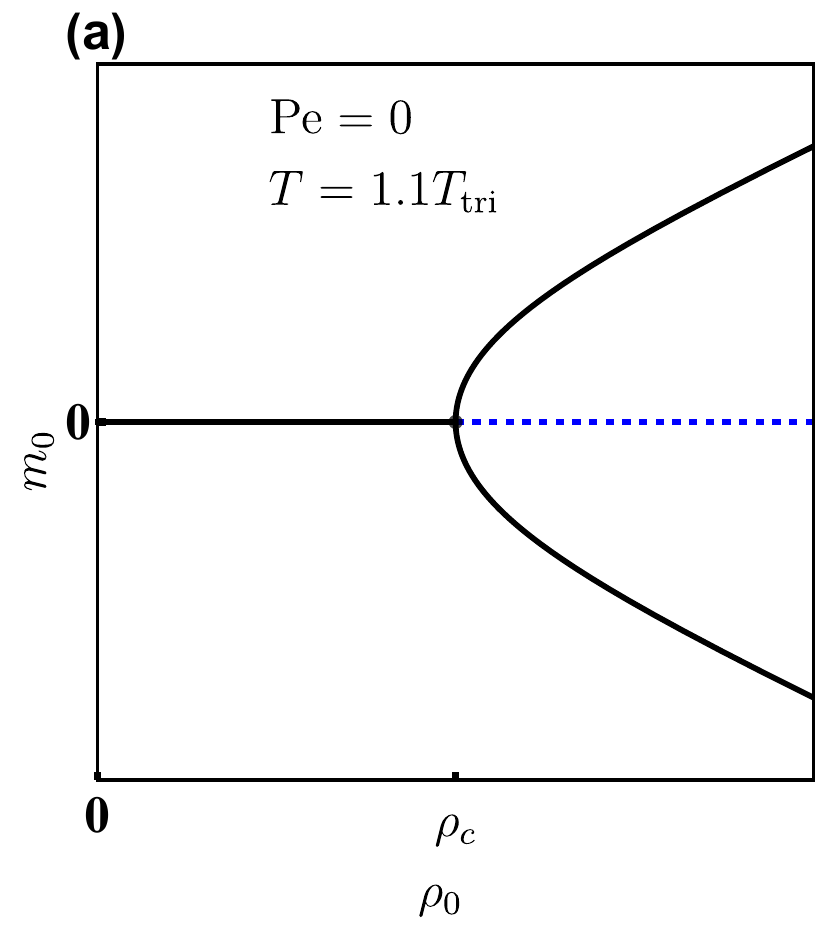}
	\includegraphics[width=.32\columnwidth]{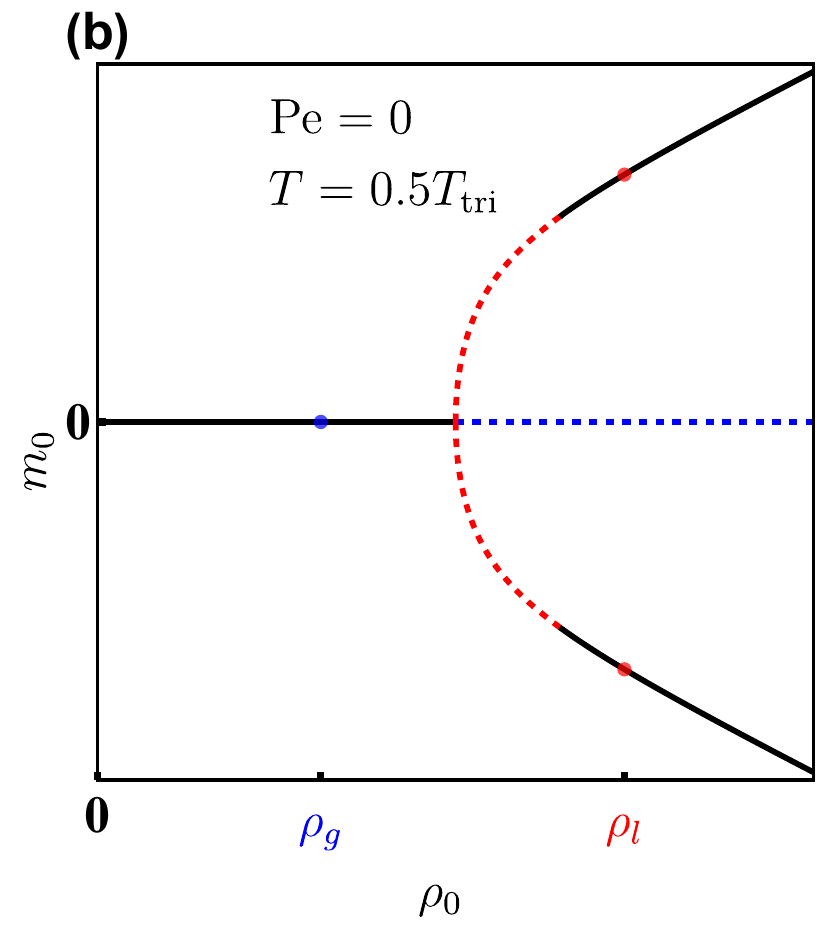}
	\caption{Equilibrium ($\text{Pe}=0$) bifurcation diagrams for (a) $T>T_{\text{tri}}$ and (b) $T<T_{\text{tri}}$, with monotonic $f$ given by~\eqref{f0} with $r=1$. Curves mark the different branches of solutions for the magnetization of homogeneous states\eqref{mstar}. Dashed blue and red lines mark instabilities of the two branches. Both originate at the density value $\rho_c$ given by the curve \eqref{inst1}. The instability of the magnetized branch (dashed red) terminates at the density value given by the density spinodal \eqref{inst2}. The region of densities enclosed between these two values is linearly unstable for both branches. Red and blue dots mark the binodal densities of the phase-separated state (P.S.), which is the globally stable state for any mean density between these points (including densities for which all homogeneous states are unstable). The black dot marks a critical point.  }
	\label{fig.bifu}
\end{figure}

\subsection{The equilibrium limit ($\text{Pe}=0$): Model C dynamics}\label{sec:mono.eq}

As pointed out in~\cite{scandolo_active_2023}, the equilibrium, zero-propulsion limit of thermodynamically consistent flocks with discrete symmetry must generically lie within the universality class of Model C~\cite{Bray2002}. In our model, this should remain true for any non-trivial choice of $f(\rho)$, while allowing different global phase behaviors even at equilibrium. Model C describes an equilibrium dynamics which couples a conserved and a non-conserved scalar order parameter, and the canonical Landau form close to criticality is given by~\cite{blume_ising_1971, roux1992sponge}
\begin{equation}\label{modelc}
	\mathcal F_{\text{cri}}=\frac{a}{2}\left(\delta\rho\right)^2+\frac{b}{4}\left(\delta\rho\right)^4+\frac{c}{2}m^2+\frac{d}{4}m^4+ e \, m^2 \delta\rho ,
\end{equation}
where higher order terms and terms purely linear in the density are omitted. Expanding our free energy in~\eqref{free} around a homogeneous neutral state $(\rho=\rho_0, m=0)$, one can derive the expressions of the different coefficients in~\eqref{modelc}, which are determined by the choice of $f\left(\rho\right)$ \footnote{The expansion also produces constant terms, a linear term $\delta\rho$, and a cubic term $\delta\rho^3$. The linear term integrates to zero due to mass conservation. The cubic term in general does not, yet it shall not affect the phase diagram, see \cite{blume_ising_1971}.}. Importantly, the coupling coefficient $e$ in~\eqref{modelc} reads
\begin{equation}\label{ccoef}
	e=\beta f'\left(\rho_0\right) ,
\end{equation}
which vanishes for a constant $f(\rho)$; in this special case, our model falls outside the Model C universality class. (As discussed in Sec.~\ref{sec.non.mon}, this also makes the nonequilibrium $\text{Pe}>0$ regime different from the usual flocking scenario.) As detailed in~\cite{roux1992sponge}, the topology of the phase diagram is set by the values of the coefficients in~\eqref{modelc}. As we have the exact form of the free energy at our disposal~\eqref{free}, we follow the procedure presented in~\cite{roux1992sponge} to map the phase diagram across the whole phase space, even beyond criticality.

We find the stable phases by minimizing the free energy~\eqref{free} under the constraint of conserved density $(1/\ell_s)\int_0^{\ell_s} \rho dx = \rho_0$. These belong to one of three possibilities: the homogeneous neutral state $(\rho=\rho_0,m=0)$ (marked in Fig. \ref{fig.phase1} by $\text{H}. m=0$), the homogeneous magnetized state $(\rho=\rho_0,m=m_0\neq0)$ (marked by $\text{H}. m\neq0$) or a phase separated state (P.S.) with spatially varying magnetization and density. In all of the stable states, since magnetization is not conserved, its value is locked to the local value of the density via the minimization of $F$ with respect to the magnetization. A non-trivial minimizer $m_0\neq0$ then emerges at $\left.\partial_{m m}\mathcal F\right\vert_{m=0}=0$ which defines the curve
 \begin{equation}\label{inst1}
T_c=f\left(\rho\right).
\end{equation} 
Generally, the relation~\eqref{inst1} only establishes linear instability of the neutral state ($m=0$). In regions that also admit phase separation, as discussed below, this curve will then mark a spinodal instability which we will term the {\em magnetization spinodal}, see dashed blue curve in Fig.~\ref{fig.phase1}(a). Only in the absence of phase separation, this curve marks a second-order transition between disordered and ordered states, see solid black curve in Fig.~\ref{fig.phase1}(a). In the ordered phase, the value $m_0\left(\rho,T\right)$ is found from $\partial_m\mathcal F=0$, which gives the magnetization in an implicit form:
\begin{equation}\label{mstar}
	m_0=\rho z(\rho,T) ,
	\quad
	\frac{\text{arctanh} z}{z}=\beta f\left(\rho\right) ,
\end{equation}
where the second equality defines the dependence of $z$ on $\rho$ and $T$. A state with magnetization~\eqref{mstar} minimizes the free energy only for homogeneous profiles. To find phase separation, one has to follow the usual procedure of a common-tangent construction over the single-variable function $\mathcal F\left[\rho,m_0\left(\rho,T\right)\right]$~\cite{roux1992sponge}. Non-convexity of this function marks linear instability against spatial density variations, and is equivalent to non-convexity of the two-variable free energy $\cal F$:
\begin{equation}\label{inst2}
	\left|\text{Hess}\left(\mathcal F\right)\right|<0 ,
\end{equation}
where $\left|\text{Hess}\left(\mathcal F\right)\right|$ is the Hessian determinant of $\mathcal F\left(\rho,m\right)$. An explicit expression for~\eqref{inst2} is presented in~\ref{ap.eq_phase}.

The relation~\eqref{inst2} defines a spinodal curve that is complementary to the magnetization spinodal~\eqref{inst1}, see dashed red line Fig.~\ref{fig.phase1}(a), which we will term the {\em density spinodal} (since instability here also involves density modulations). The two spinodal curves meet at a tricritical point, as detailed in~\ref{ap.eq_phase}, whose location  obeys for the density $\rho_{\text{tri}}$
\begin{equation}\label{cond0}
\left[\rho_{\text{tri}}\frac{f'\left(\rho_{\text{tri}}\right)}{f\left(\rho_{\text{tri}}\right)}\right]^2=\frac{2}{3} .
\end{equation}
The region enclosed between the two spinodals \eqref{inst1} and \eqref{inst2} defines the region of spinodal decomposition. This is shown by the bifurcation diagram Fig. \ref{fig.bifu} which marks instabilities along a constant temperature cross section of Fig.\ref{fig.phase1}(a). At $T>T_{\text{tri}}$ (Fig. \ref{fig.bifu} (a)) the neutral branch $m=0$ loses stability at the density given by \eqref{inst1} while the entire branch of magnetized states $m_0\neq0$, prescribed by the nontrivial solution to \eqref{mstar}, is stable. In contrast, at $T<T_{\text{tri}}$ (Fig. \ref{fig.bifu} (b)), the magnetized branch has an unstable segment which terminates at the spinodal density \eqref{inst2}. Thus, any homogeneous state with density within the region enclosed between the two spinodals \eqref{inst1} and \eqref{inst2} is linearly unstable.

Any monotonic $f(\rho)$ that vanishes at a finite density while saturating at infinity will have at least one root of~\eqref{cond0}, yielding a phase diagram like that in Fig.~\ref{fig.phase1}(a). At $T>T_c\left(\rho_{\text{tri}}\right)$, the curve~\eqref{inst1} marks a line of critical points separating the homogeneous disordered ($m=0$) and ordered ($m\neq0$) phases. At $T<T_c\left(\rho_{\text{tri}}\right)$ the transition to the ordered phase becomes discontinuous with a region of coexistence between an ordered liquid and a disordered gas. Within the miscibility gap, the relation~\eqref{inst1} marks the magnetization spinodal, and the complementary density spinodal given by~\eqref{inst2} lies within the symmetry-broken phase. The location of the binodal curves follows from a common tangent construction, as detailed in~\ref{ap.eq_phase}. Note that, although coexisting bulk phases are found within this procedure, properly resolving the interfaces would necessitate retaining sub-leading gradient terms, which quantify surface tension in the free energy. Note also that Model C can exhibit additional phase diagram topologies, such as critical points and double critical end points~\cite{roux1992sponge}, under appropriate tuning of $f(\rho)$. These are absent for the curves~\eqref{f0}, but could arise for other sufficiently elaborate choices, including strictly monotonic curves.


\subsection{The nonequilibrium regime ($\rm{Pe}>0$): Discontinuous flocking transition}\label{sec:mono.noneq}

The phase diagram for the monotonic $f(\rho)$ obeying~\eqref{f0} at finite $\text{Pe}=0.5$ is shown in Fig.~\ref{fig.phase1}(b). The tricritical point shifts to infinite density, turning the second-order transition line into a first order coexistence region (miscibility gap) along its entire length. The same phase behavior is observed for AIM and its recent variants~\cite{Solon2013, scandolo_active_2023}. Within the miscibility gap, the phase separated state becomes a traveling band (T.B.), as shown in Fig.~\ref{fig.phase1}(c). Indeed, at non-vanishing self-propulsion, any magnetized bulk phase attains a non-vanishing density flux given by $\text{Pe}\,m$, according to~\eqref{d2}. Then, simple flux balance across the interface implies that it must propagate with velocity given by
\begin{equation}\label{v}
	V=\text{Pe}\frac{\Delta m}{\Delta \rho} ,
\end{equation}
where $\Delta \rho$ and $\Delta m$ are the density and magnetization difference between phases. Moreover, we have found numerically that the propagation velocity is bounded below as $|V|\geq\text{Pe}$, for all models that we have examined.

In the absence of a free energy minimization principle, there is no simple recipe to establish the binodals in the nonequilibrium case. Throughout this paper, the nonequilibrium phase diagrams were found by numerically integrating the hydrodynamics (\ref{d1}-\ref{d2}) using a finite difference scheme for space and time. Then binodals are identified as the plateaus of the traveling density profiles as shown in Fig.\ref{fig.phase1}(c). For more details see \ref{ap.num}.

To establish more broadly the generic topology of the phase diagram in Fig.~\ref{fig.phase1}(b), we analyze the linear instability of the dynamics. First, the magnetization spinodal line~\eqref{inst1} remains unchanged even at finite $\text{Pe}>0$, since it describes instabilities against \textit{homogeneous} variations. As the active terms in the dynamics~(\ref{d1}-\ref{d2}) only couple to spatial flux derivatives, they do not affect this instability. In contrast, the complementary density spinodal ~\eqref{inst2} is changed. As we show in~\ref{ap.non_eq_phase} the instability in~\eqref{inst2} is now modified into the following criterion
\begin{equation}\label{instp}
	\left|\text{Hess}\left({\cal F}\right)\right|<\frac{\text{Pe}^2}{2M\left(\rho_0,m\right)\rho_0}\left[\left(\frac{\partial_{\rho m}\mathcal F}{\partial_{m m}\mathcal F}\right)^2-1\right] ,
\end{equation}
where the mobility $M$ is given in~\eqref{mob}, and the magnetization $m$ is to be evaluated at $m=m_0(\rho,T)$ in~\eqref{mstar}.

Based on other flocking models, it was argued that the flocking transition should generically be discontinuous everywhere~\cite{Chate2020}. In our model, this holds true if the conditions in~\eqref{inst1} and~\eqref{instp} predict a finite spinodal gap at any $\text{Pe}>0$. It is useful to notice that, at the magnetized state $m=m_0(\rho,T)$ in~\eqref{mstar}, we get $\partial_{\rho m}  \mathcal F/\partial_{mm} \mathcal F=-\partial m_0/\partial\rho$, so that the density spinodal~\eqref{instp} becomes
\begin{equation}\label{instp1}
\left|\text{Hess}\left({\cal F}\right)\right|<\frac{\text{Pe}^2}{2M\left(\rho_0,m\right)\rho_0}\left[\left(\frac{\partial m_0}{\partial\rho}\right)^2-1\right] .
\end{equation}
Moreover, close to the magnetization spinodal~\eqref{inst1}, we have $m_0\sim\sqrt{T_c\left(\rho\right)-T}$ with $T_c=f(\rho)$, yielding
\begin{equation}\label{eq:dm}
\left.\frac{\partial m_0}{\partial\rho}\right\vert_{\rho_0}\sim \frac{T_c'\left(\rho_0\right)}{\sqrt{T_c\left(\rho_0\right)-T}}. 
\end{equation}
Given that~\eqref{eq:dm} diverges close to the magnetization spinodal~\eqref{inst1}, a finite neighborhood of the magnetization spinodal remains unstable at arbitrary small $\text{Pe}$ no matter where one is on the phase diagram. Therefore, the non-perturbative effect of activity on the phase diagram can be attributed to the singularity of the order disorder transition of the equilibrium limit, as encoded in $m_0$. Remarkably, however, this argument fails for any non-monotonic $f(\rho)$ at its maximum $f'(\rho_0)=0$. As discussed next in Sec.~\ref{sec.non.mon}, the case of non-monotonic $f$ actually leads to novel phase behaviors and a diagram very different from Fig.~\ref{fig.phase1}(b).


\section{Phase diagram for non-monotonic alignment strength}\label{sec.non.mon}

\begin{figure}[b]
	\centering
	\includegraphics[width=.32\columnwidth]{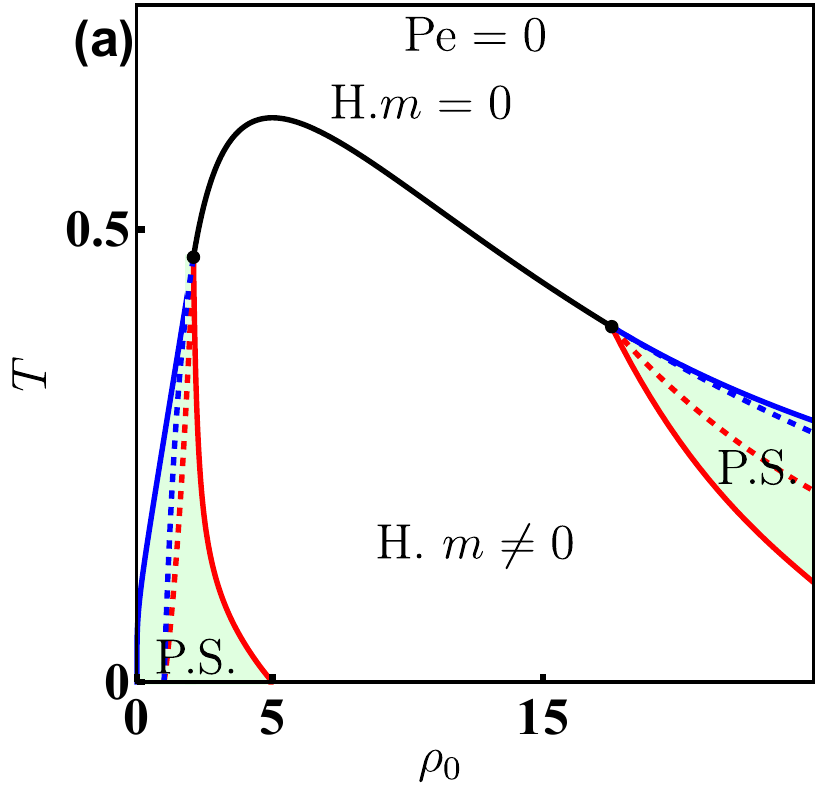}
	\includegraphics[width=.32\columnwidth]{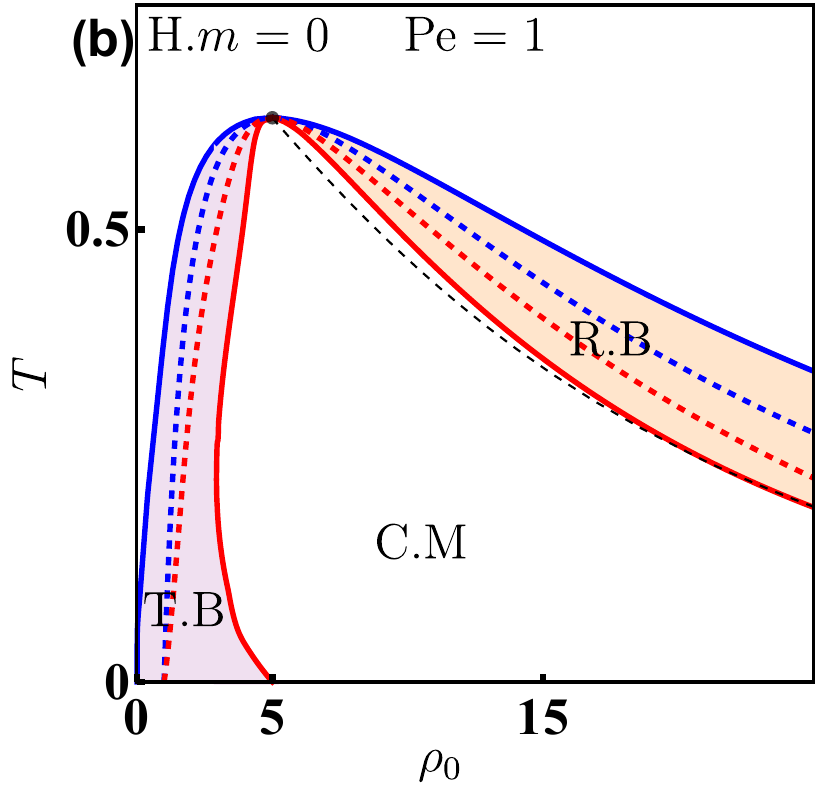}	
	\includegraphics[width=.315\columnwidth]{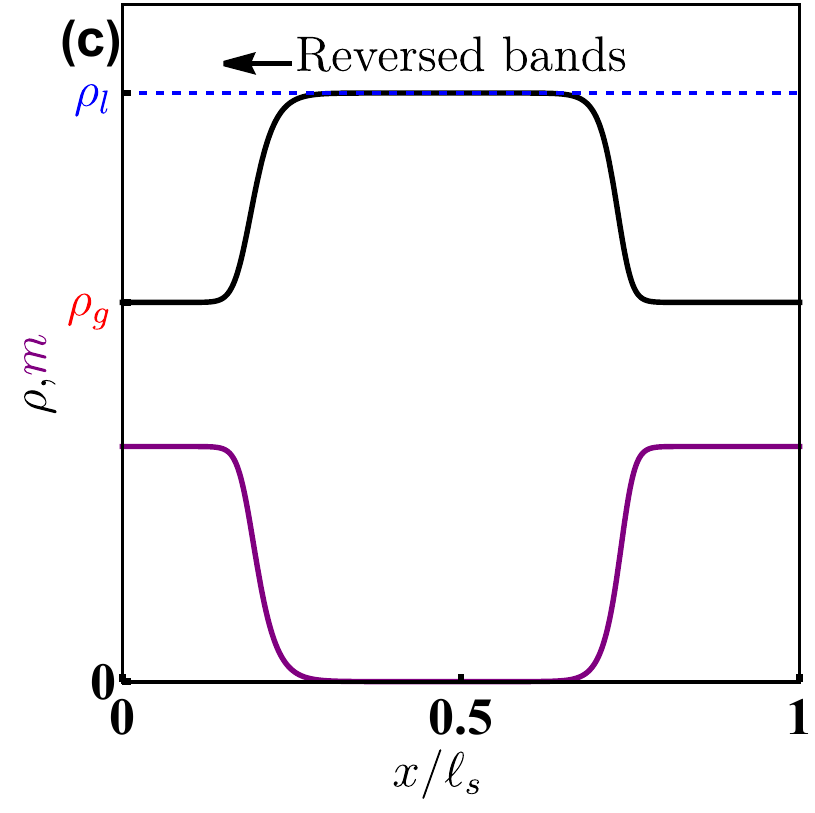}
	\caption{Phase diagrams for (a)~the equilibrium case ($\text{Pe}=0$) and (b)~the nonequilibrium regime ($\text{Pe}=1$), with non-monotonic $f$ given by~\eqref{gen} with $\alpha=1/20$. The colored dashed and solid curves are spinodals and binodals, respectively. The black solid curve is a critical line. The black dots mark tricritical points. The black dashed line marks the maximal magnetization curve~\eqref{rhobarimp}. An `azeotropic point' emerges in the nonequilibrium regime.
	(c)~Reversed band (R.B.) profiles for $\text{Pe}=1$, $\rho_0\simeq15.62$, and $T\simeq0.43$.}
	\label{fig.phase2}
\end{figure}

Non-monotonic alignment strength functions $f(\rho)$ are likely to arise physically in systems where alignment is restricted by visual obstruction, as was recently proposed to model natural bird flocking \cite{pearce_role_2014}. Non-monotonicity opens the door to novel phase behaviors, not encountered for the monotonic case in Sec.~\ref{sec:mono} above. In equilibrium ($\text{Pe}=0$), when $f(\rho)$ features a single maximum, one typically finds two roots for~\eqref{cond0} marking a pair of tricritical points. This scenario leads to two distinct miscibility gaps, as shown in Fig.~\ref{fig.phase2}(a). The miscibility gap to the left of the maximum has a disordered low-density phase coexisting with an ordered higher-density phase. The miscibility gap to the right of the maximum has a disordered high-density phase coexisting with an ordered lower-density phase.

In the rest of this Section, we consider how self-propulsion ($\text{Pe}>0$) affects this equilibrium picture. The regime of non-monotonic $f$ is more elaborate compared with the monotonic case in Sec.~\ref{sec:mono}. In particular, we distinguish two qualitatively different phase behaviors, depending on the curvature of $f(\rho)$ close to its maximum. For illustration purposes, we use particular examples in a specific family of non monotonic-curves:
\begin{equation}\label{gen}
	f\left(\rho\right)=\left(1-\frac{1}{\rho}\right)e^{-\alpha\rho} ,
\end{equation} 
with $\alpha>0$. In what follows, we show that the topology of the phase diagram is controlled by local properties of $f$ close to the maximum, so that the exponential cutoff at large densities is immaterial for our purposes.


\subsection{Azeotropic point in the nonequilibrium case ($\rm{Pe}>0$)}\label{sec:azeo}

For mildly curved functions $f(\rho)$, we find numerically that the tricritical points collide at the maximum of $f$ at any finite $\text{Pe}$, see Fig.~\ref{fig.phase2}(b). As for the case of monotonic $f(\rho)$ in Sec.~\ref{sec:mono}, the effect of self-propulsion on the phase diagram is non-perturbative: the equilibrium critical line expands almost everywhere into a finite miscibility gap, except at a single point which remains critical. Thus, in contrast with the case of monotonic $f(\rho)$, there does remain a critical point on the phase diagram at finite $\text{Pe}$. Therefore, a non-monotonic $f(\rho)$ can allow for a continuous transition from homogeneous disorder (H., $m=0$) to collective motion (C.M.).

For equilibrium systems showing a phase-diagram topology equivalent to Fig.~\ref{fig.phase2}(b), the critical point is called an azeotropic point~\cite{koynova_gel-state_1987}, so that we also adopt the same term here. This point marks the pairwise meeting of two branches of first order lines, with miscibility gaps that display distinct behavior. The left gap features the usual traveling bands (T.B.) as in Fig.~\ref{fig.phase1}(c). The right gap features reversed bands (R.B.), shown in Fig.~\ref{fig.phase2}(c). These bands propagate in the direction opposite to the bulk magnetization. Indeed, in the corresponding miscibility gap, the coexistence is between a magnetized lower-density phase and a paramagnetic high-density phase. Then, according to the flux balance condition~\eqref{v}, it follows that the phase boundaries propagate in the reverse direction from usual.

Our scenario shows that the flocking transition need not {\em necessarily} be discontinuous, at variance with a widely held view among those studying flocking~\cite{Chate2020}, although it remains an exceptional case. In order to rule out some continuous transition scenarios, note that the disordered (H., $m=0$) and the ordered phases (H., $m\neq0$ at $\text{Pe}=0$ or C.M. at $\text{Pe}\neq0$) entail different symmetries. Thus, it is not possible to pass between these pure phases without crossing a (symmetry-breaking) phase transition of some kind. This argument precludes the standard scenario of liquid-liquid phase separation, where a single pair of binodals terminate at a critical point.  Indeed, one sees from the equilibrium phase diagrams in Figs.~\ref{fig.phase1}(a) and \ref{fig.phase2}(a) that the binodals terminate at tricritical points, and any path between the pure phases must cross a phase transition line.  In the nonequilibrium phase diagrams of Figs.~\ref{fig.phase1}(b) and \ref{fig.phase2}(b), one observes discontinuous transitions as well as an azeotropic point, all of which are consistent with the relevant symmetry principles.

\subsection{Counterpropagating bands in the nonequilibrium case ($\rm{Pe}>0$)}

\begin{figure}
	\centering
	\includegraphics[width=.32\columnwidth]{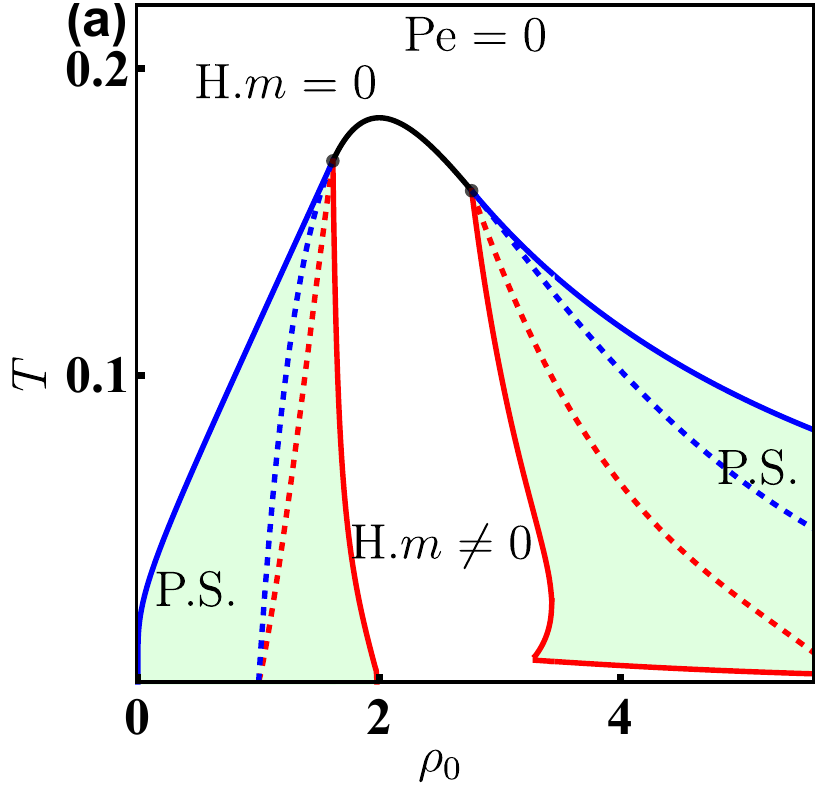}
	\includegraphics[width=.345\columnwidth]{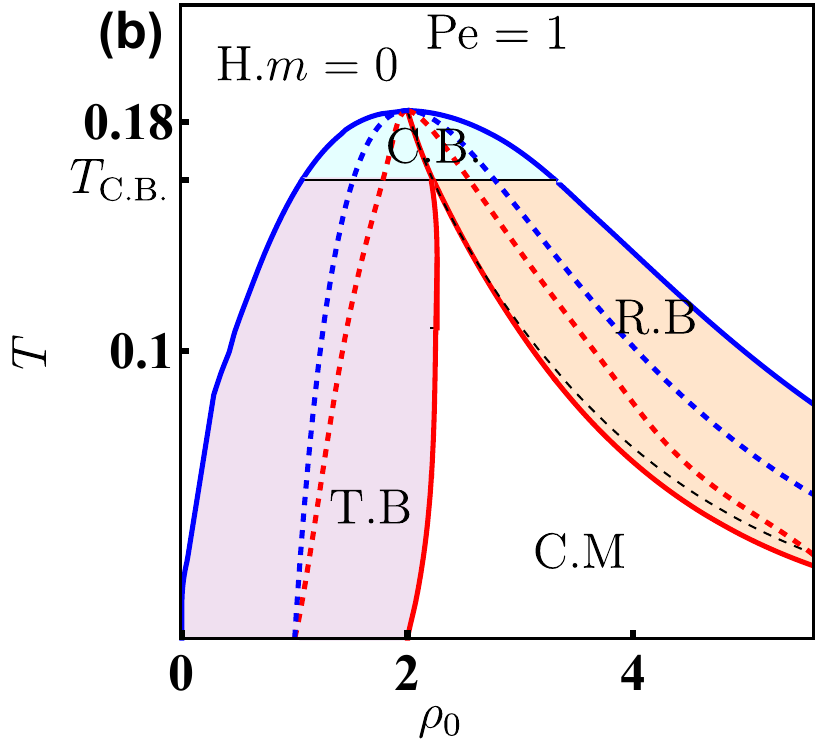}	
	\includegraphics[width=.315\columnwidth]{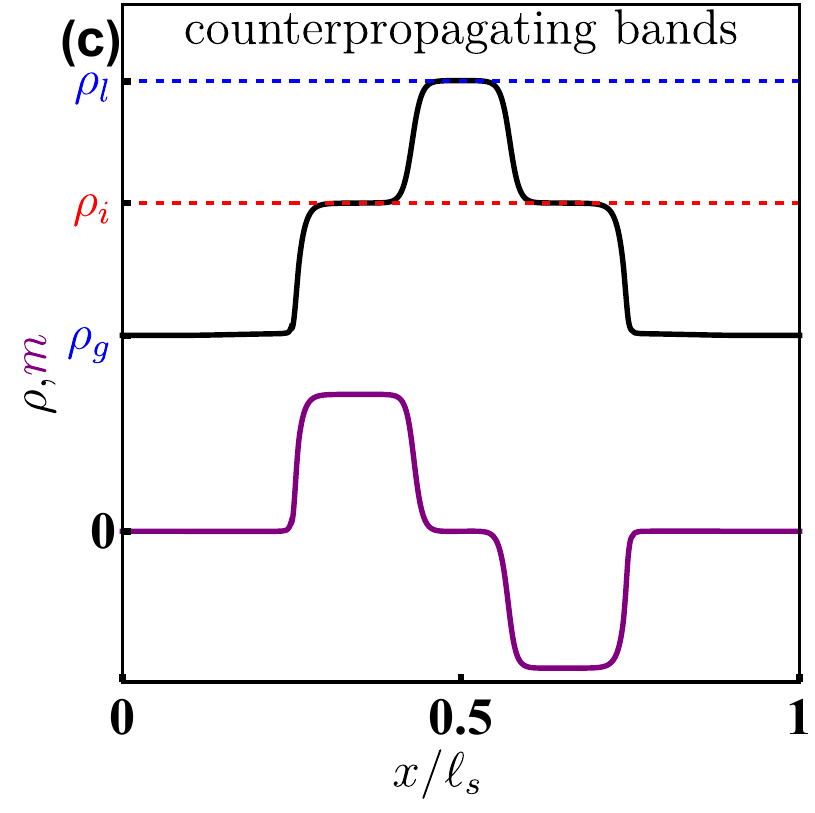}\\
	\caption{Phase diagrams for (a)~the equilibrium case ($\text{Pe}=0$) and (b)~the nonequilibrium regime ($\text{Pe}=1$), with non-monotonic $f$ given by~\eqref{gen} with $\alpha=1/2$. The colored dashed and solid curves are spinodals and binodals, respectively. The black solid curve is a critical line. The black dots mark tricritical points. The black dashed line marks the maximal magnetization curve~\eqref{rhobarimp}. A C.B. phase emerges in the nonequilibrium regime.
	(c)~The counterpropagating bands profiles for $\text{Pe}=1$, $\rho_0\simeq1.8$, and $T\simeq0.17$.
	}
	\label{fig.phase3}
\end{figure}

When $f(\rho)$ is sufficiently curved, two tricritical points in the equilibrium limit are closer to the maximum of $f(\rho)$, see Fig.~\ref{fig.phase3}(a). Here, we observe a different behavior at finite $\text{Pe}$ compared with Sec.~\ref{sec:azeo}. Instead of the azeotropic point reported in Fig.~\ref{fig.phase2}(b), there now appears a region with counterpropagating bands (C.B.), as shown in Fig.~\ref{fig.phase3}(b). This happens when the two magnetised binodals (red solid lines) merge, at a finite temperature $T_{\text{C.B.}}$.  For temperatures between $T_{\text{C.B.}}$ and $T_{\text{max}}=\text{ max} \left[f(\rho)\right]$ , the system enters the C.B. phase (see Fig.~\ref{fig.phase3}(c)), a strongly fluctuating dynamical steady state whose properties can be understood in terms of the partial restoration of a broken symmetry, as we discuss next.

Recall that within the symmetry-broken T.B. and R.B. phases, the steady state consists of a single magnetized band, which travels through the system.  The magnetization may be either positive or negative: this spontaneous breaking of spin-reversal symmetry is familiar from the Ising model.  In turn, this symmetry breaking drives a particle current, which causes the band to move either left or right. This additionally breaks the spatial reflection symmetry of the model.  Due to the periodic boundary conditions of the system, the resulting steady state is also time-periodic, with period $\ell_s/V$.

\begin{figure}
	\centering
	\includegraphics[width=.19\columnwidth]{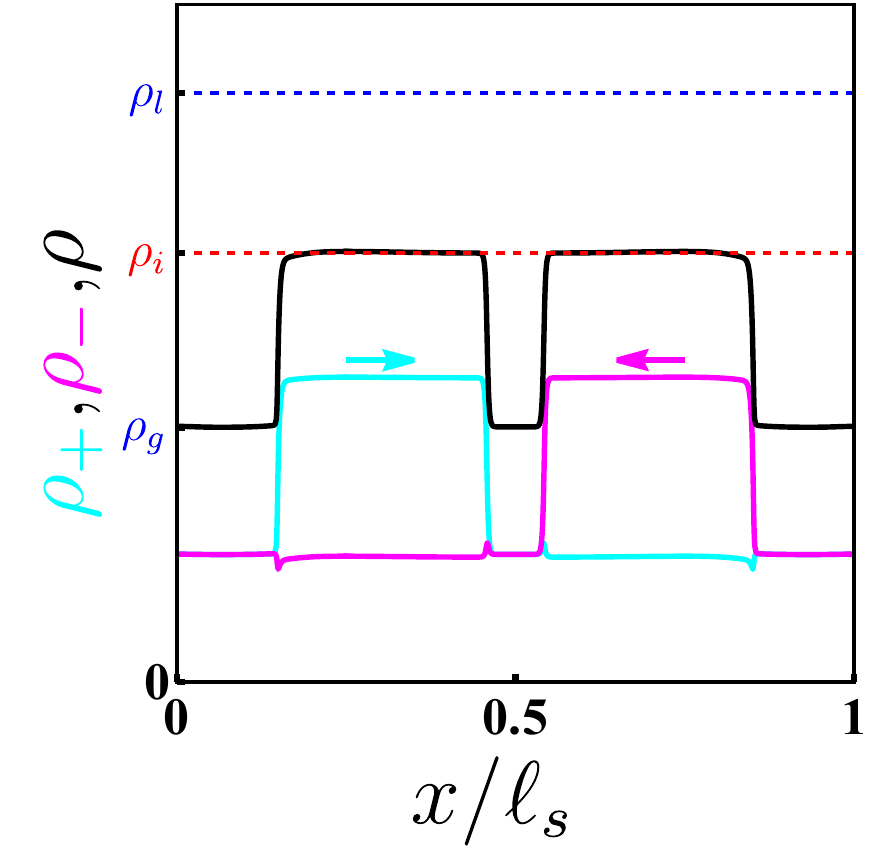}
	\includegraphics[width=.19\columnwidth]{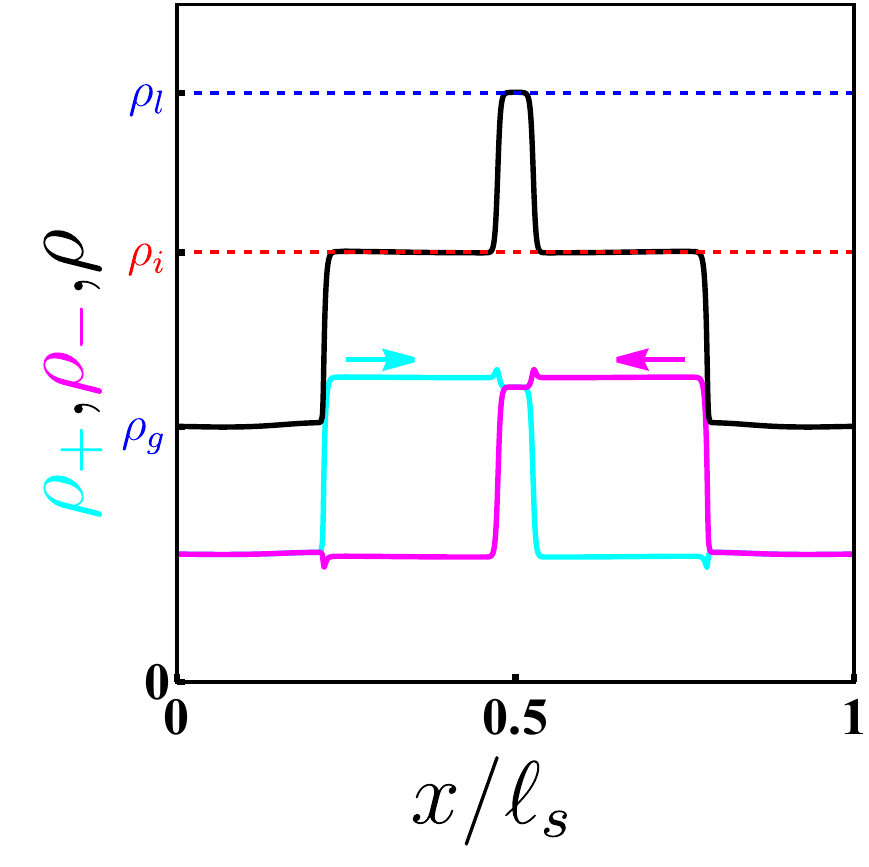}
	\includegraphics[width=.19\columnwidth]{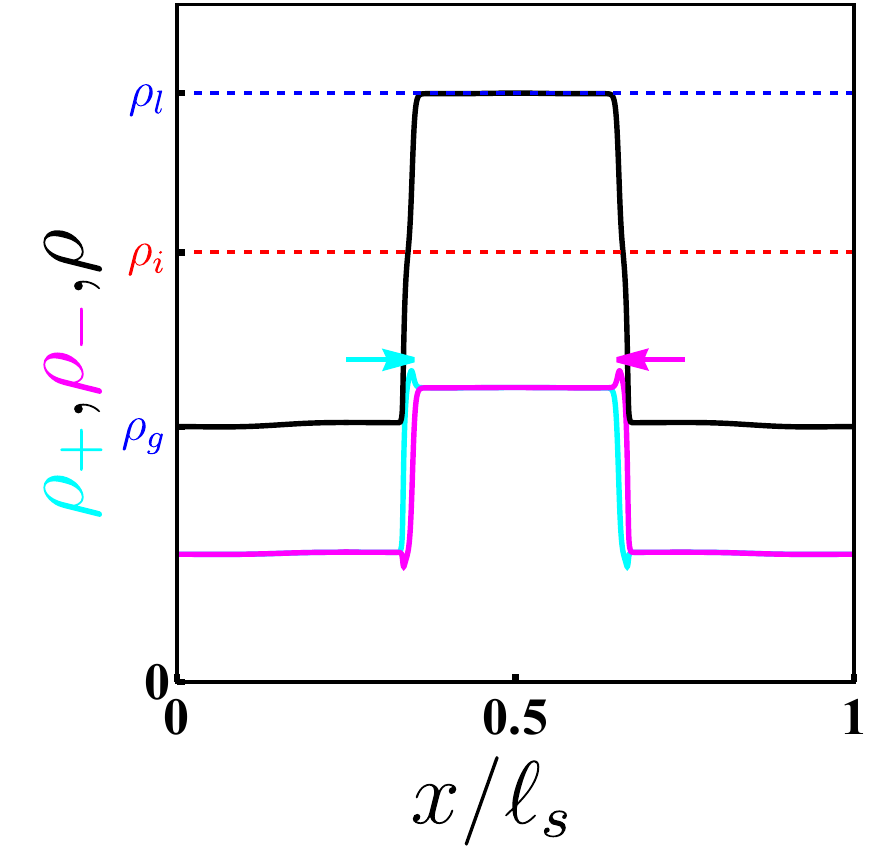}
	\includegraphics[width=.19\columnwidth]{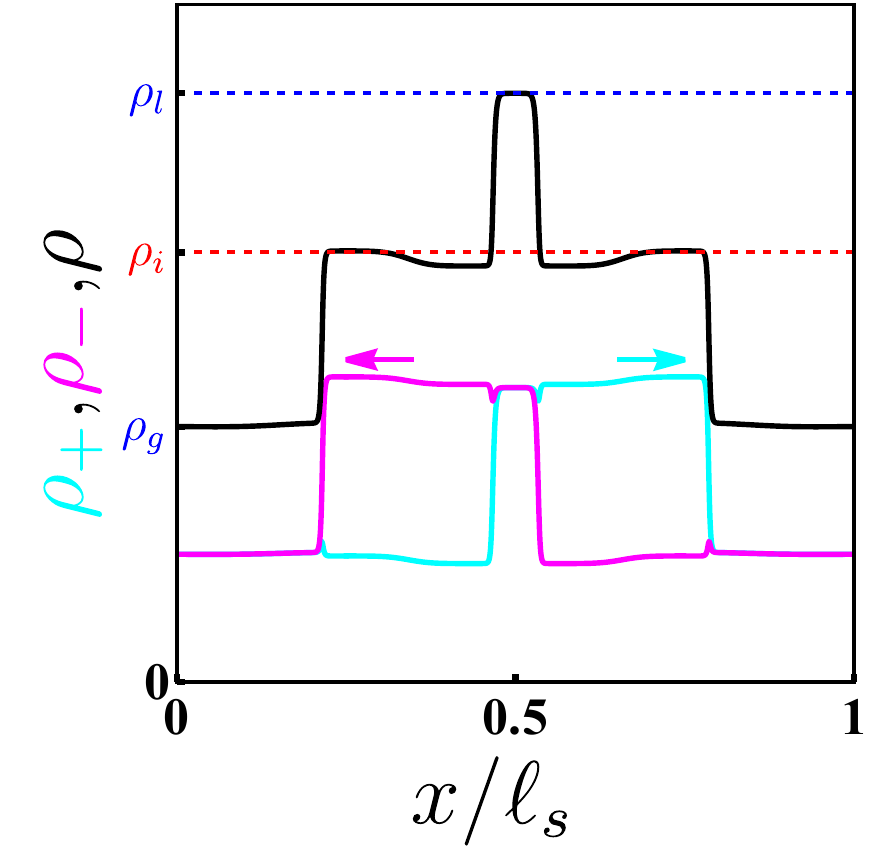}
	\includegraphics[width=.19\columnwidth]{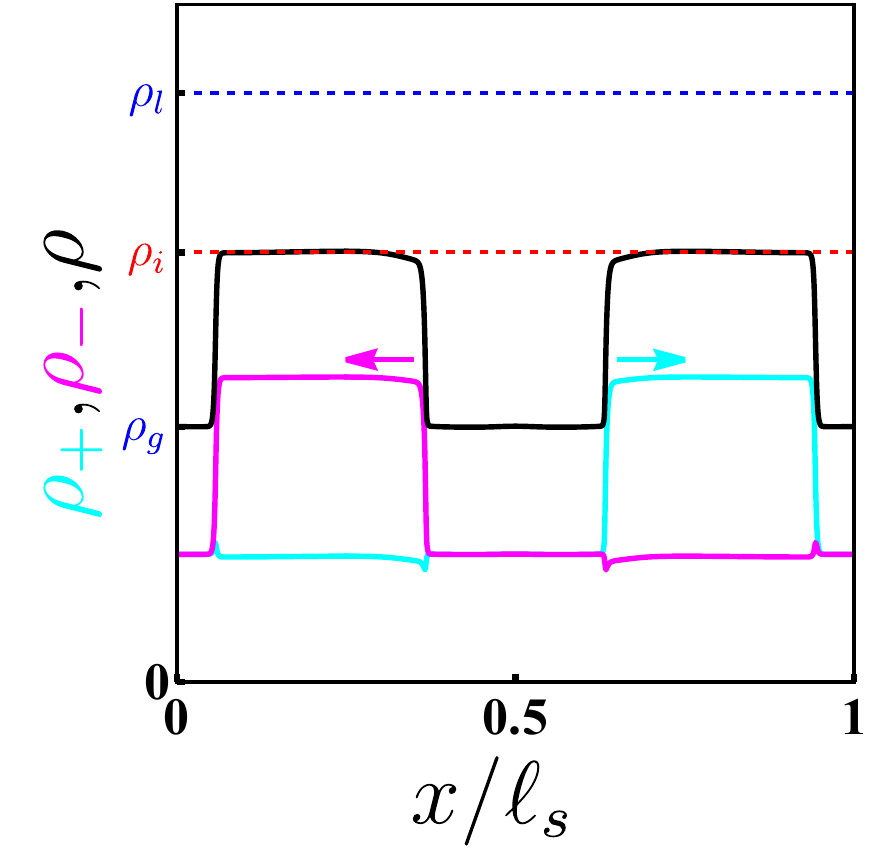}
	\vskip.1cm
	\includegraphics[width=.19\columnwidth]{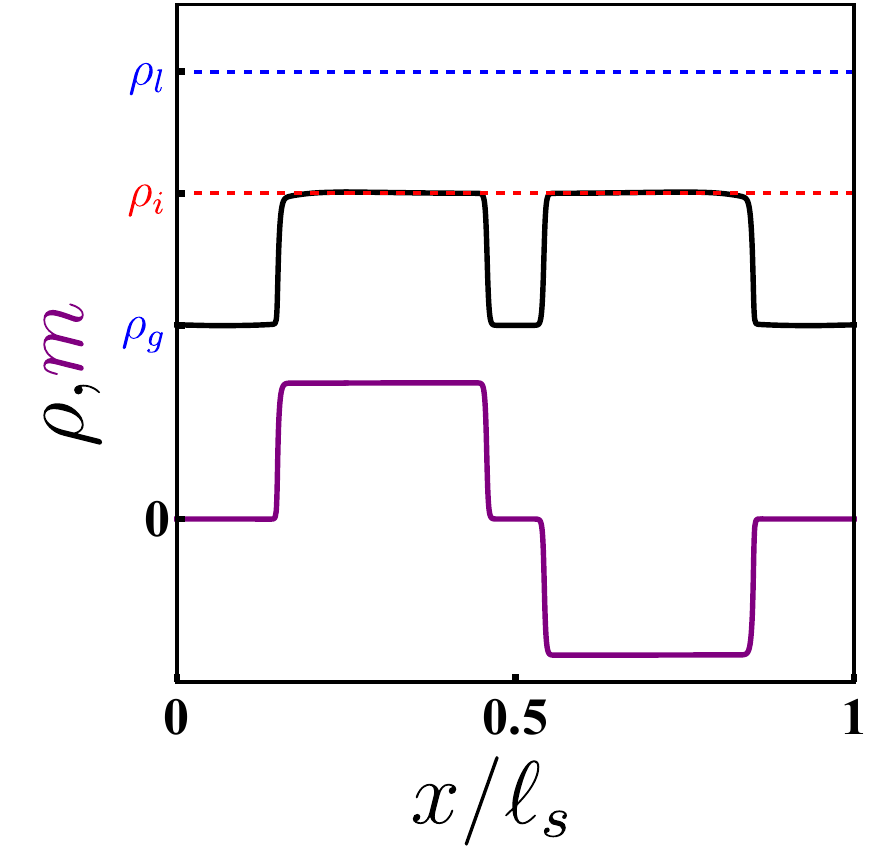}
	\includegraphics[width=.19\columnwidth]{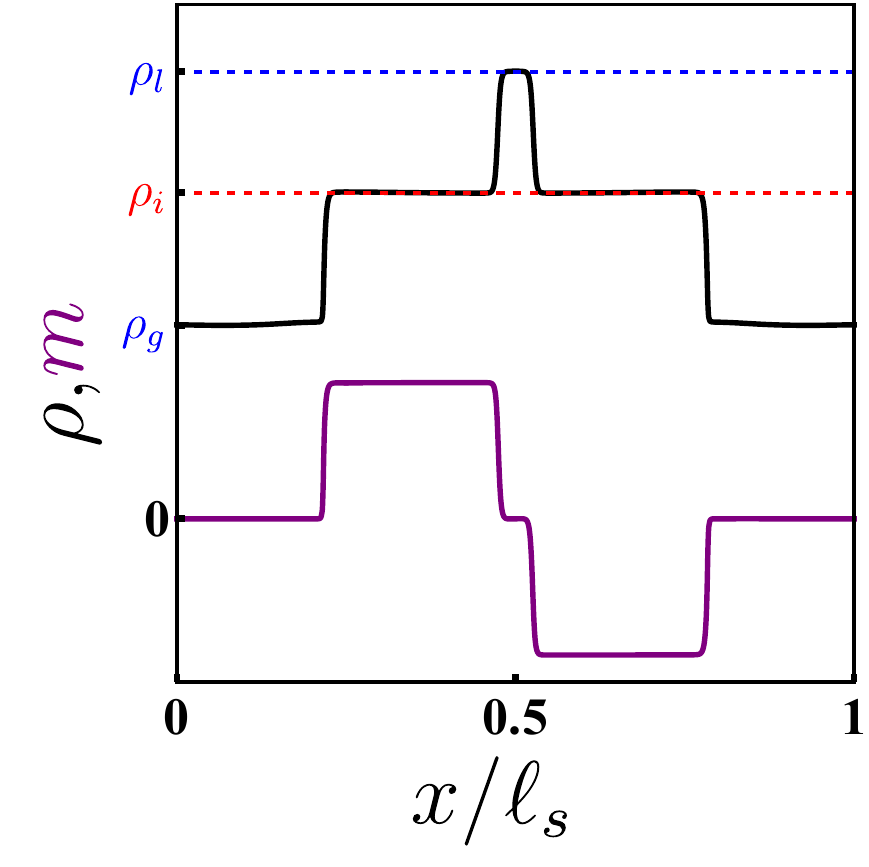}
	\includegraphics[width=.19\columnwidth]{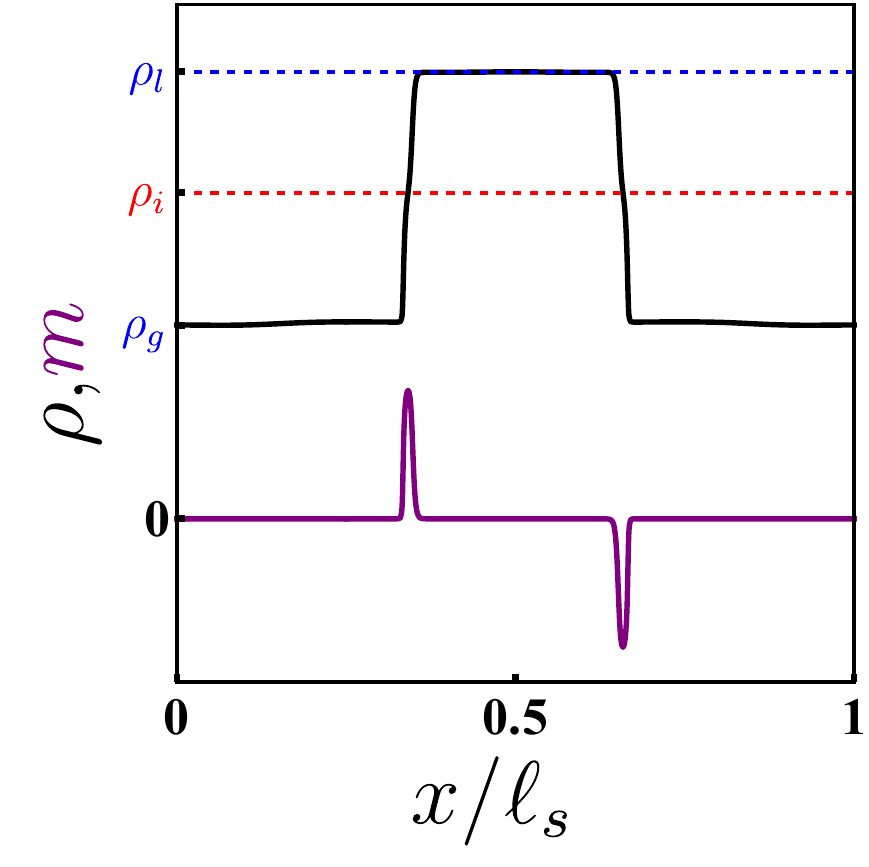}
	\includegraphics[width=.19\columnwidth]{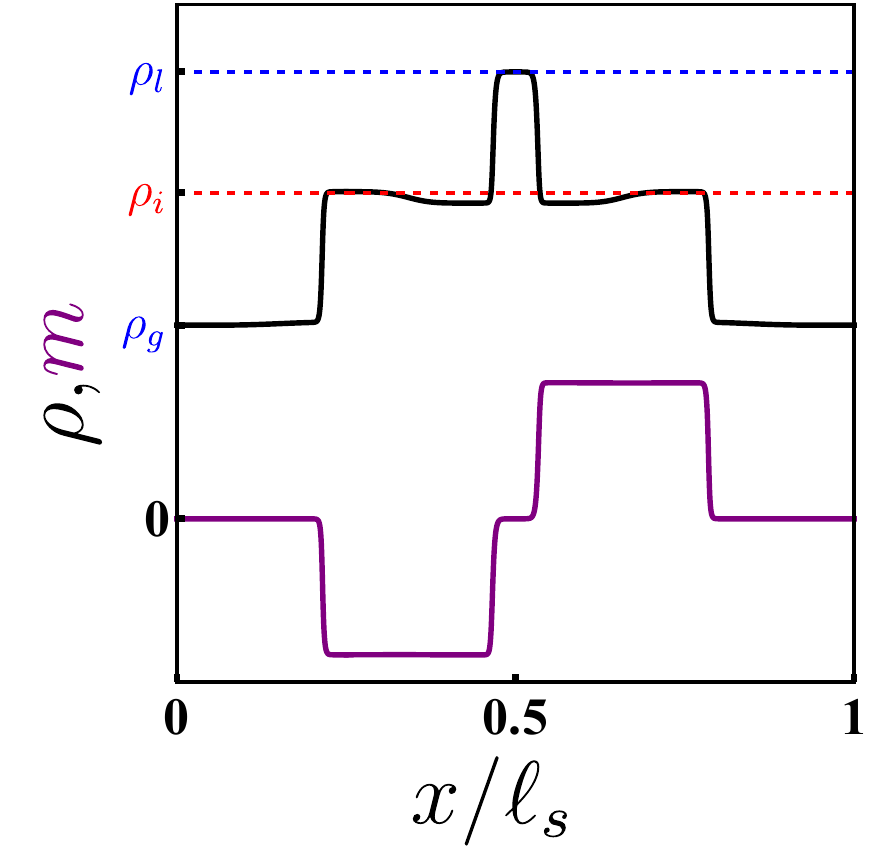}
	\includegraphics[width=.19\columnwidth]{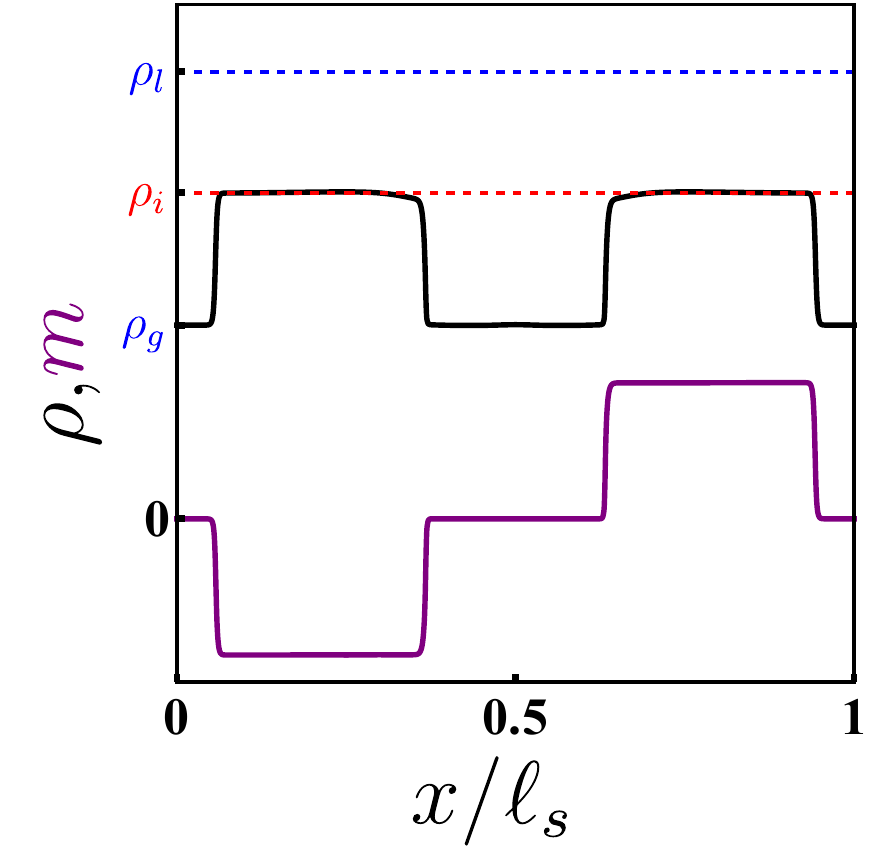}\\
	\includegraphics[width=.5\columnwidth]{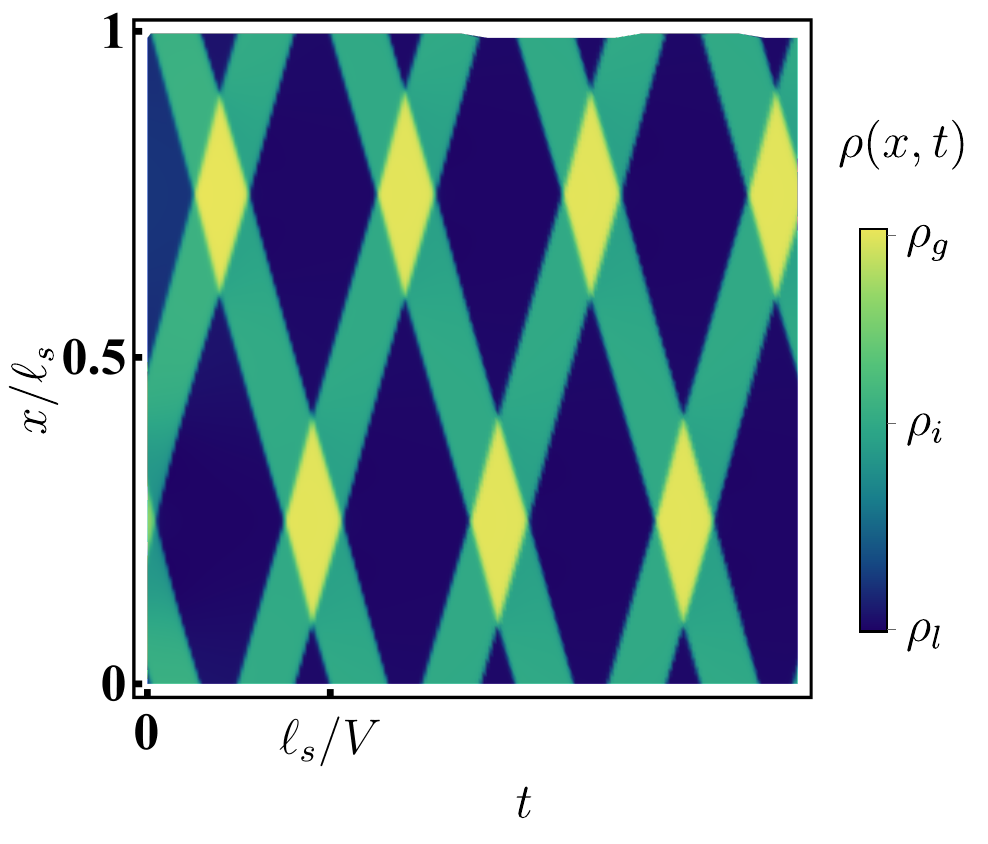}
	\caption{Time evolution of the counterpropagating bands for $\text{Pe}=1$, $\rho_0\simeq1.8$, and $T\simeq0.17$. (Top)~The counterpropagating bands of the $\rho_{\pm}$ fields, in cyan and magenta, barely scatter upon encounter. (Middle)~The corresponding density $\rho=\rho_++\rho_-$ and magnetization $m=\rho_+-\rho_-$ display an oscillating dynamical state with volume fraction periodically varying in time. (Bottom)~the corresponding space time plot for the density profile.}
	\label{fig.pro.tp}
\end{figure}

\begin{figure}
	\centering
	\includegraphics[width=.3\linewidth]{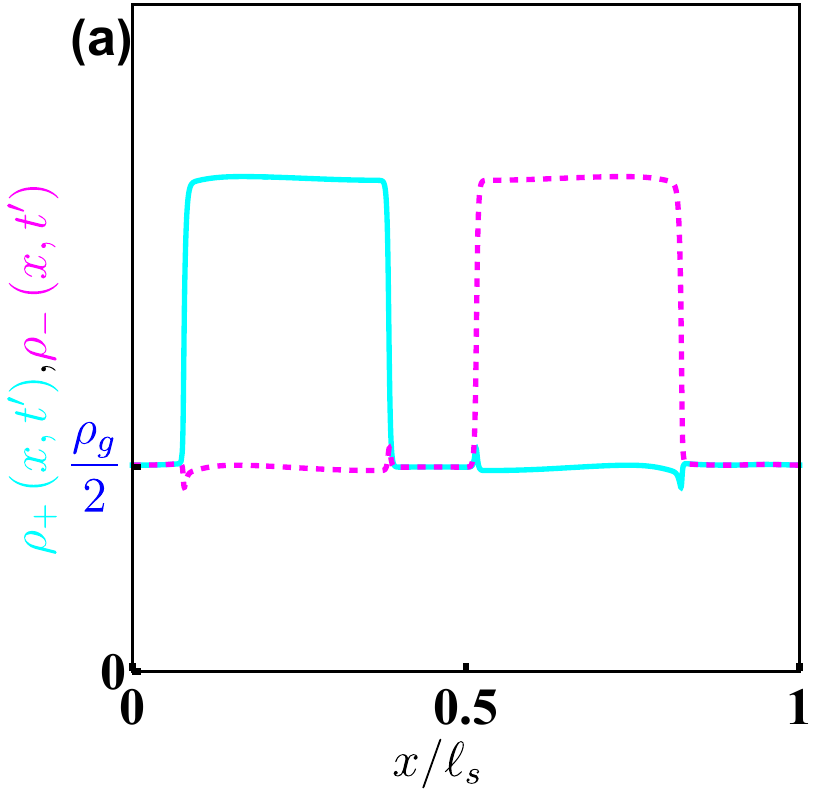}
	\includegraphics[width=.3\linewidth]{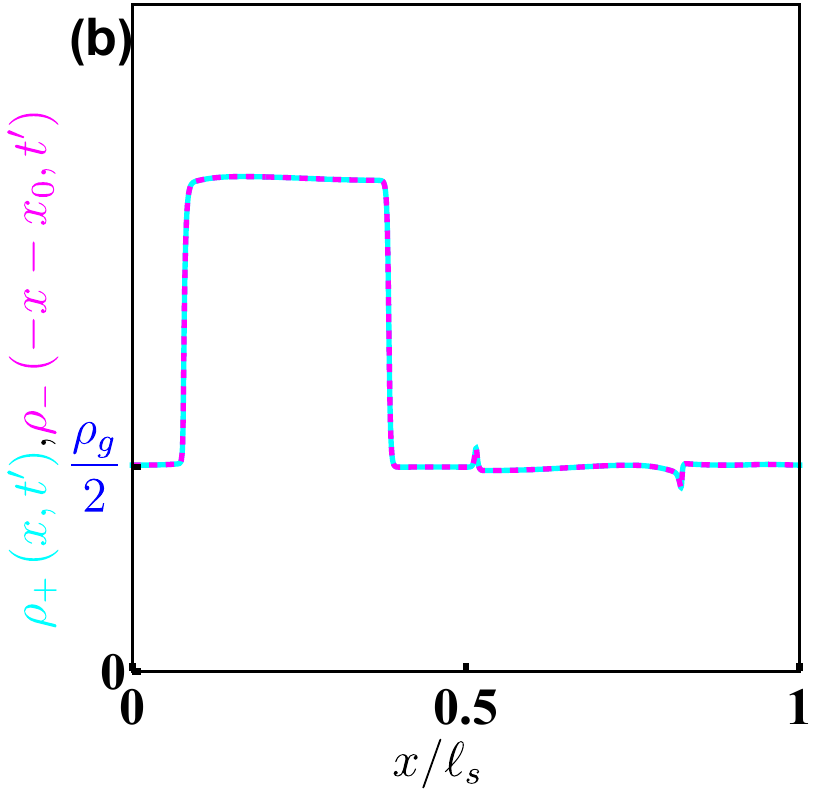}
	\includegraphics[width=.33\linewidth]{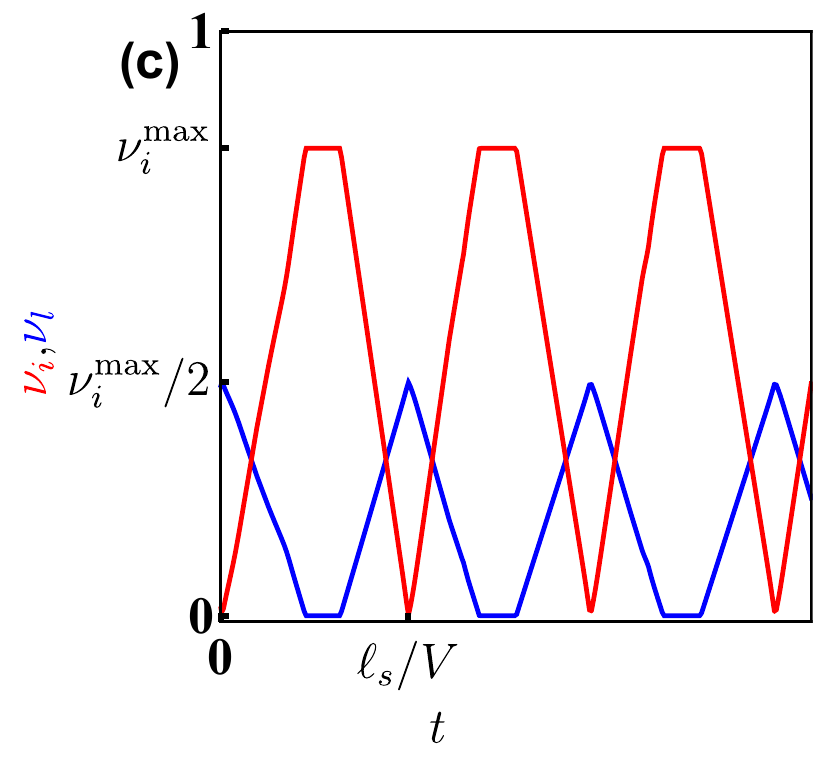}
	\caption{ (a-b)~The $\pm$ counterpropagating bands profiles are perfectly anti-symmetric: $\rho_+\simeq\bar{\rho}\left(x-Vt\right)$ and $\rho_-\simeq\bar{\rho}\left(-x-Vt\right)$ coincide when reflected and superimposed.
		(c)~The periodic evolution of the phase volumes of the intermediate and liquid phases $\nu_i$ and $\nu_l$ respectively, for parameter values in Fig.~\ref{fig.pro.tp}. The complementary gas phase volume (not shown) is simply $\nu_g=1-\nu_i-\nu_l$. }
	\label{fig.frac}
\end{figure}

In the C.B. phase, both left- and right-moving bands are present at the same time, as shown in Fig.~\ref{fig.pro.tp}. Since they propagate in opposite directions, the two bands regularly meet each other. It turns out that they interact weakly (we recall that the number of particles on each site is unbounded, in contrast with exclusion processes).  The resulting situation is most simply represented in terms of the density of $+$ and $-$ particles
\begin{equation}
\rho_\pm\left(x,t\right)=\frac{\rho\left(x,t\right)\pm m\left(x,t\right)}{2} .
\end{equation}
As shown in Figs.~\ref{fig.frac}(a-b), the space-time dependence of these profiles is well-approximated by $\rho_+\simeq\bar{\rho}\left(x-Vt\right)$ and $\rho_-\simeq\bar{\rho}\left(-x-Vt\right)$. In this case, the density $\rho(x,t)$ is still time-periodic, with behaviour invariant under spatial reflection, so that it is analogous to a standing-wave solution, while the T.B. and R.B. are analogous to travelling-wave solutions. These analogues of standing waves and travelling waves are strongly anharmonic, as expected since the governing equations are non-linear. In fact, $\bar{\rho}$ is close to a top-hat profile, as it was in the T.B./R.B. phases: there are extended bulk regions at densities $(\rho_1,\rho_2)$, separated by interfaces whose widths are $\mathcal O\left(1\right)\ll\ell_s$. Plotting the density profile at generic times, as shown in Fig.~\ref{fig.pro.tp}, one typically finds up to four distinct regions whose densities are the following. Outside of either band the density is that of the dilute binodal $\rho_g=2 \rho_1$. Inside both bands, the density is that of the dense binodal $\rho_l=2 \rho_2$. Finally, inside one band and outside the other, the density is that of the intermediate binodal $\rho_i=\rho_1+\rho_2$.  Hence the intermediate binodal is always halfway between the extreme binodals, while their difference equals twice the magnetization $m_0$:
\begin{equation}\label{inter}
\rho_g\left(T\right)=\rho_{i}\left(T\right)-m_0\left(T\right) , 
\quad
\rho_l\left(T\right)=\rho_{i}\left(T\right)+m_0\left(T\right).
\end{equation} 
An important consequence of~\eqref{inter} is that the propagation velocity~\eqref{v} attains its limiting minimal value $|V|=\text{Pe}$. The magnetization in~\eqref{inter} marks the bulk magnetization of the intermediate binodal $m_0=m_0\left(\rho_i,T\right)$, as given explicitly by~\eqref{mstar}. Therefore, both binodals $\rho_{g,l}$ in~\eqref{inter} are fully determined by the value of the intermediate binodal $\rho_i$. In practice, we find empirically that the binodal $\rho_i$ coincides with the maximal magnetization curve $\rho_{\text{max}}$, defined by $\partial_\rho m_0(\rho_{\text{max}},T)=0$, as shown in dashed black line in Fig.~\ref{fig.phase3}(b). We get an implicit representation of $\rho_{\text{max}}(T)$ by differentiating~\eqref{mstar} with respect to $\rho$, yielding
\begin{equation}\label{rhobarimp}
	\frac{z}{\left(1-z^2\right)\arctanh z}=1-\frac{\rho_{\text{max}} f'\left(\rho_{\text{max}}\right)}{f\left(\rho_{\text{max}}\right)},
	\quad T=\frac{z}{\arctanh z} f\left(\rho_{\text{max}}\right).
\end{equation}
Overall, Equations~(\ref{inter}-\ref{rhobarimp}) specify all three binodals of the C.B. phase in closed analytic form.

We now address the phase volumes of the bulk phases, defined as the fraction of space that each phase occupies, and the corresponding highly dynamical nature of the C.B. phase state. Since the profiles of $\rho_+$ and $\rho_-$ are symmetric, each one of them carries exactly half of the total mass $L\rho_0/2$. Besides, given that $\pm$ bands counter-propagate at constant speed, their overlap region varies periodically with time. We report in Fig.~\ref{fig.frac}(c) the corresponding dynamics of the phase volumes for a mean density below the intermediate binodal ($\rho_0<\rho_i$). The maximal phase volume of the intermediate phase $\nu_i^{max}$ is set by the usual lever rule between the intermediate and gas binodals: $\nu_i^{max}=\left(\rho_i-\rho_0\right)/\left(\rho_i-\rho_g\right)$. Then, the maximal phase volume for the liquid phase occurs at perfect overlap of the $\pm$ profiles and is exactly $\nu_i^{max}/2$, as shown in Fig.~\ref{fig.frac}(c).

Interestingly, one can formulate a necessary condition for the emergence of the C.B. phase state. Indeed, the liquid and gas binodals must lie outside the spinodals $\rho_g^s$ and $\rho_l^s$:
\begin{equation}\label{cond_tri}
\rho_g\left(T\right)<\rho_g^s\left(T\right) < \rho_l^s\left(T\right)<\rho_l\left(T\right) .
\end{equation}
The condition~\eqref{cond_tri} must hold across the C.B. phase. In particular, it should hold in the vicinity of the maximum of $f$, where the C.B. phase first emerges. Approximating $f(\rho)$ by a parabola in this region, the spinodal densities $\rho_{g,l}^s$ defined in~\eqref{inst1} are symmetrically positioned around the maximum $\rho^*$:
\begin{equation}\label{delta}
\rho_g^s\simeq\rho^*-\Delta_s, 
\quad
\rho_l^s\simeq\rho^*+\Delta_s,
\quad
\Delta_s = \sqrt{\frac{2\epsilon f(\rho^*)}{\left|f''\left(\rho^*\right)\right|}} ,
\quad
\epsilon = \frac{f(\rho^*)-T}{ f(\rho^*)}\ll1 .
\end{equation}
Performing a similar expansion for the maximal magnetization curve in~\eqref{rhobarimp} yields
\begin{equation}\label{eq:m*}
	\rho_{\text{max}}\left(T\right)\simeq\rho^* ,
	\quad
	m_0\left(\rho_{\text{max}}, T\right)\simeq\rho^*\sqrt{3\epsilon} .
\end{equation}
Substituting the expression~\eqref{eq:m*} in~\eqref{inter}, and recalling that $\rho_i=\rho_{\text{max}}$ in the C.B. phase, we find that the liquid and gas binodals are also symmetrically positioned around the maximum:
\begin{equation}\label{deltab}
\rho_g\simeq\rho^*-\Delta_b ,
\quad
\rho_l\simeq\rho^*+\Delta_b ,
\quad
\Delta_b=\sqrt{3\epsilon}\rho^* .
\end{equation}  
Finally, the condition~\eqref{cond_tri} for existence of the C.B. phase is equivalent to the ordering $\Delta_b>\Delta_s$ of the binodal and spinodal gaps. Using the expressions for these gaps, respectively in~\eqref{delta} and~\eqref{deltab}, we arrive at the following condition for the curvature near the maximum:
\begin{equation}\label{cond_curve}
\frac{\rho^{*2}\left|f''\left(\rho^*\right)\right|}{f\left(\rho^*\right)}>\frac{2}{3} .
\end{equation}
Remarkably, the condition~\eqref{cond_curve} is independent of the P\'eclet number, which corroborates that the C.B. phase appears as a singular perturbation of the underlying equilibrium phase diagram at any finite activity, no matter how small.


\section{Discussion}

Our results show that the density-dependence of the aligning interactions can play an unexpectedly major role in flocking models, yielding a richer flocking phenomenology than any previously reported. These results are found by choosing a thermodynamically consistent approach which entails a proper equilibrium limit, pertaining to the universality class of Model C~\cite{Bray2002, scandolo_active_2023}, at vanishing self-propulsion. Such rich phenomenology should not be restricted to thermodynamically consistent models provided an appropriate choice of density dependent interactions. However, we find that thermodynamically consistent models are best suited for anticipating and understanding these phenomena. In particular, the spinodal condition \eqref{inst1} which is given in terms of the equilibrium free energy, helps rationalize why a tricritical point in the equilbirium limit (Fig. \ref{fig.phase1} (a)) `runs away' to infinite density in the presence of self propulsion (Fig. \ref{fig.phase1} (b)), recovering the standard scenario of discontinuous flocking~\cite{Chate2020}. Moreover, for non-monotonic alignment $f$, one has that the pair of tricritical points that exist in the equilibrium limit (Fig. \ref{fig.phase2} (a)) must collide at the maximum of $f$ in the presence of self propulsion, giving way to more exotic phase topologies which could involve either an azeotropic point (Fig. \ref{fig.phase2} (b)) or a region of counterpropagating bands phase (Fig. \ref{fig.phase3} (b)).  Remarkably, in contrast with standard flocking models~\cite{Chate2020}, the flocking transition is continuous in the presence of an azeotropic point, while remaining discontinuous elsewhere on the phase diagram.

Our models employ a diffusive scaling of the microscopic rates to allow for an analytically exact derivation of a hydrodynamic description. However, the phenomenology is expected to persist for models without such scaling. Indeed, although the hydrodynamic description of the AIM employs a similar diffusive scaling, microscopic simulations show that the predicted flocking transition phenomenology holds without such scaling, see \cite{solon_flocking_2015}. The mesoscopic interaction range $\Delta x$ used here enables these phenomena to appear in our simple one-dimensional setting. We expect that in higher dimensions much of the phenomenology would survive also for locally interacting models.

Given the rich topology of these phase diagrams, it would be interesting to examine how the entropy production rate (EPR) varies across the various phase transitions that we have unveiled. In thermal systems~\cite{Lebowitz1999, seifert_stochastic_2012}, EPR measures both the breakdown of time-reversal symmetry and the amount of energy dissipated by nonequilibrium currents. For flocking systems, irreversibility measures have been provided in both microscopic models~\cite{Shim2016, Spinney2018, Giardina2022, Tu2022} and field theories~\cite{Ramaswamy2018, Borthne2020}. Yet, these measures should coincide with a quantification of dissipation only for thermodynamically consistent models~\cite{Markovich2021, Falasco2023, Proesmans2024}. In that respect, our approach opens the unprecedented opportunity to properly study how the dissipation varies across the flocking transition. For instance, it could lead to a spatial resolution of dissipation, thus uncovering the profile of energy dissipation associated with travelling bands.


\section*{Acknowledgements}

The authors acknowledge useful discussions with M. Esposito, G. Falasco, K. Proesmans, and A. Tanaji Mohite. This research was funded in part by the Luxembourg National Research Fund (FNR), grant reference 14389168. For the purpose of open access, \'EF has applied a Creative Commons Attribution 4.0 International (CC BY 4.0) license to any Author Accepted Manuscript version arising from this submission.


\appendix

\section{Deriving hydrodynamic equations}\label{ap.hydro}

The microscopic dynamics described in Sec.~\ref{sec:def} consists of biased diffusive hops along with reaction dynamics. To derive the corresponding hydrodynamics, we follow the recipe outlined in~\cite{bodineau_current_2010-1}, which has already been implemented in other models involving active lattice gases~\cite{kourbane-houssene_exact_2018-1, agranov_exact_2021, agranov_entropy_2022, agranov_macroscopic_2023}. A necessary property for the derivation is that the underlying microscopic dynamics is dominated by symmetric diffusive hops, which have the fastest transition rate.

First, we express the Hamiltonian~\eqref{h} in terms of the hydrodynamic fields~\eqref{den}. Assuming that these fields are slowly varying over the macroscopic scale (which will be verified {\em a posteriori}), then one has, to leading order at large $L$,
\begin{equation}\label{h2}
H = \frac{L}{\ell_s}\int_0^{\ell_s}dx\mathcal 	H \left[\rho\left(x,t\right),m\left(x,t\right)\right] , 
\end{equation}  
with the Hamiltonian density
\begin{equation}
\mathcal H\left(\rho,m\right)=-\frac{m^2}{2\rho}f\left(\rho\right) .
\end{equation}	
When a $+$ particle at position $i=Lx_i/\ell_s$ hops to the right, the field variations follow as
\begin{equation}\label{eq:rho}
\rho\to \rho +\frac{\ell_s}{L}\delta\left(x-x_{i+1}\right)-\frac{\ell_s}{L}\delta\left(x-x_i\right)=\rho-\frac{\ell_s^2}{L^2}\partial_x\delta\left(x-x_i\right)+o\left(L^{-2}\right) .
\end{equation}
Similarly, we have
\begin{equation}\label{eq:m}
m\to m -\frac{\ell_s^2}{L^2}\partial_x\delta\left(x-x_i\right)+o\left(L^{-2}\right) ,
\end{equation}
and analogous expressions hold for the other jumps. The expression for the field variations under a flip of $+\to-$ is simply given by
\begin{equation}
\rho\to \rho ,
\quad
m\to m -2\frac{\ell_s}{L}\delta\left(x-x_i\right)+o\left(L^{-1}\right) .
\end{equation}
Substituting the expressions~(\ref{eq:rho}-\ref{eq:m}) into~\eqref{h2}, we find that the jump of a $+$ particle to the right leads to the Hamiltonian increment
\begin{equation}\label{eq:dh1}
\Delta H=\frac{\ell_s}{L}\partial_x\left(\frac{\delta H}{\delta \rho}+\frac{\delta H}{\delta m}\right) ,
\end{equation}
with similar expressions hold for the other jumps. The Hamiltonian increment under $+\to-$ flipping is simply
\begin{equation}\label{eq:dh2}
\Delta H=-2\frac{\delta H}{\delta m} .
\end{equation}
Substituting~(\ref{eq:dh1}-\ref{eq:dh2}) in the definition of the rates of Sec.~\ref{sec:def}, we find that these rates are functions of the local fields to leading order:
\begin{equation}\label{u}
\begin{aligned}
\text{rate}_+(i\rightarrow i+1) &=D_0\left[1+\frac{1}{L}\left(\frac{\lambda}{D_0}-\frac{\beta}{2}\ell_s\partial_x\left.\left(\frac{\delta H}{\delta\rho}+\frac{\delta H}{\delta m}\right)\right\vert_{\rho(x,t),m(x,t)}\right)\right] ,
\\
\text{rate}_+(i\rightarrow i-1) &=D_0\left[1+\frac{\beta\ell_s}{2L}\left.\left(\frac{\delta H}{\delta\rho}+\frac{\delta H}{\delta m}\right)\right\vert_{\rho(x,t),m(x,t)}\right] ,
\\
\text{rate}_-(i\rightarrow i+1) &=D_0\left[1-\frac{\beta\ell_s}{2L}\left.\left(\frac{\delta H}{\delta\rho}-\frac{\delta H}{\delta m}\right)\right\vert_{\rho(x,t),m(x,t)}\right] ,
\\
\text{rate}_-(i\rightarrow i-1)&=D_0\left[1+\frac{1}{L}\left(\frac{\lambda}{D_0}+\frac{\beta}{2}\ell_s\partial_x\left.\left(\frac{\delta H}{\delta\rho}-\frac{\delta H}{\delta m}\right)\right\vert_{\rho(x,t),m(x,t)}\right)\right] ,
\\
\text{rate}(+\rightarrow -) &= \frac{\gamma}{L^2}e^{\beta\frac{\delta H}{\delta m}} .
\end{aligned}
\end{equation}
Importantly, the leading order terms are symmetric hops with rate $D_0$. These dominate over both the asymmetric $\mathcal O\left(L^{-1}\right)$ jump rate, and over the slower $\mathcal O\left(L^{-2}\right)$ tumbling rate defined in Sec.~\ref{sec:def}. This timescale separation guarantees that the measure is controlled by the simple symmetric hop rules corresponding to noninteracting random walkers~\cite{bodineau_current_2010-1}. It is then straightforward to write down the coupled hydrodynamics of the $\rho_+$ and $\rho_-$ fields
\begin{equation}\label{da1}
\begin{bmatrix}
\partial_t{\rho_+}\\
\partial_t{\rho_-}
\end{bmatrix}=-\partial_x\begin{bmatrix}
J_{\rho_+}\\
J_{\rho_-}
\end{bmatrix}+
\begin{bmatrix}
-K\\
K
\end{bmatrix} .
\end{equation}
The hydrodynamic fluxes are found by coarse-graining, using the hydrodynamic coordinates $x=\ell_si/L$ and $t=\gamma\hat{t}/L^2$, with $\hat{t}$ the time variable of the microscopic dynamics:
\begin{equation}\label{j}
\begin{aligned}
J_{\rho_+} &=-\partial_x\rho_+-\beta\rho_+\partial_x\left(\frac{\delta H}{\delta\rho}+\frac{\delta H}{\delta m}\right)+\text{Pe}\rho_+ ,
\\
J_{\rho_-} &=-\partial_x\rho_--\beta\rho_-\partial_x\left(\frac{\delta H}{\delta\rho}-\frac{\delta H}{\delta m}\right)-\text{Pe}\rho_- ,
\end{aligned}
\end{equation} 
and, following the same steps, the tumbling rate reads
\begin{equation}
K=me^{\beta\frac{\delta H}{\delta m}} .
\end{equation}
Note that, when sending $\beta\to0$ in~\eqref{j}, one recovers the fluxes of AIM~\cite{Solon2013}. Finally, one obtains the expressions~(\ref{d2}-\ref{free}) in terms of $\rho$ and $m$ with a few steps of algebra. 
Interestingly, in the equilibrium limit ($\text{Pe}=0$), the free energy $ F$ serves as a Lyapunov function:
\begin{equation}\label{lyap}
\frac{d  F}{dt}=-\frac{L}{\ell_s}\int_0^{\ell_s} dx\left\{\frac{1}{2} \partial_x \begin{bmatrix}
\frac{\delta  F}{\delta\rho}\\
\frac{\delta  F}{\delta m}\end{bmatrix}^\dag\mathbb C(\rho,m)\partial_x\begin{bmatrix}
\frac{\delta  F}{\delta\rho}\\
\frac{\delta  F}{\delta m}\end{bmatrix}+2M\frac{\delta  F}{\delta m}\sinh\left(\frac{\delta F}{\delta m}\right)\right\}
\leq 0 ,
\end{equation}
where we have used that $M>0$ and $\mathbb C$ is positive semi-definite. The case $d{ F}/dt=0$ is achieved in steady state. In the presence of noise, this case corresponds to the system eventually relaxing towards the minimum $\delta  F/\delta \rho=\delta  F/\delta m=0$.
Notice, however, that the well-posedness of this hydrodynamic description must break in the equilibrium limit $\text{Pe}=0$ for any non-convex free energy $\mathcal F$~\eqref{free}. It is cured by retaining sub-leading interfacial terms $\mathcal O\left(L^{-2}|\nabla\rho|^2,L^{-2}|\nabla m|^2\right)$.


\section{Spinodals and binodals in the equilibrium limit}\label{ap.eq_phase}

As mentioned in Sec.~\ref{sec:mono.eq}, the density spinodal~\eqref{inst2}, which is complementary to the magnetization one~\eqref{inst1}, marks the boundary of the convex region of the single-variable function $\mathcal F\left[\rho,m_0\left(\rho,T\right)\right]$~\cite{roux1992sponge}. Indeed, this boundary is given by
\begin{equation}\label{inst1a}
\left.\frac{\partial^2\mathcal F}{\partial \rho^2}\right\vert_{\rho_0,m_0\left(\rho,T\right)}+\left.\frac{\partial^2\mathcal F}{\partial \rho\partial m}\right\vert_{\rho_0,m_0\left(\rho,T\right)}\left.\frac{\partial m_0\left(\rho,T\right)}{\partial \rho}\right\vert_{\rho_0}=0 ,
\end{equation}
where we used the fact that, in the ordered state $m_0\left(\rho_0,T\right)$, we have
\begin{equation}\label{fm}
\left.\frac{\partial\mathcal F}{\partial m}\right\vert_{\rho_0,m_0\left(\rho,T\right)}=0 .
\end{equation}
Differentiating~\eqref{fm} with respect to $\rho$ yields
\begin{equation}\label{inst0a}
\left.\frac{\partial m_0\left(\rho,T\right)}{\partial \rho}\right\vert_{\rho_0}=-\left.\left(\frac{\partial^2\mathcal F}{\partial \rho\partial m} \bigg/\frac{\partial^2\mathcal F}{\partial ^2 m} \right)\right\vert_{\rho_0,m_0\left(\rho_0,T\right)} .
\end{equation}
Plugging the relation~\eqref{inst0a} into~\eqref{inst1a}, it follows that finding the density spinodal is equivalent to locating the onset of local non-convexity of the two-variable free energy:
\begin{equation}\label{inst2a}
\left|\text{Hess}\left(\mathcal F\right)\right|_{\rho_0,m_0\left(\rho_0,T\right)}=0 .
\end{equation} 
The density spinodal can be expressed in an implicit parametric form, since all partial derivatives entering~\eqref{inst2a} are explicit functions of $\rho$ and $m$. Indeed, we can use the parametric representation~\eqref{mstar}
\begin{equation}\label{tx}
z = \frac{m_0}{\rho_0} , 
\quad
T\left(z,\rho_0\right)=\frac{z}{\text{arctanh}(z)}f\left(\rho_0\right) ,
\end{equation}
to explicitly write~\eqref{inst2a} as
\begin{equation}\label{spin}
\left(\frac{1}{1 - z^2} - \frac{\text{arctanh}(z)}{z}\right)\left[1-\frac{x\text{arctanh}(z)}{2}\frac{(d/d\rho)\left(\rho^2 f'\left(\rho\right)\right)}{f\left(\rho\right)}\right] = \text{arctanh}^2(z)\left(\frac{\rho f'\left(\rho\right)}{f\left(\rho\right)}\right)^2 ,
\end{equation}
which gives $z=z(\rho_0)$ in an implicit form, from which follows the curve $T\left(\rho_0\right)$ for the density spinodal. This procedure allows us to determine the corresponding curves in Figs.~\ref{fig.phase1}(a),~\ref{fig.phase2}(a) and~\ref{fig.phase3}(a).

The binodals are found by the common-tangent construction over the single-variable free energy $\mathcal F\left[\rho,m_0\left(\rho_0,T\right)\right]$, which can result in several possible phase-diagram topologies~\cite{roux1992sponge}. The binodals correspond to the coexistence between dense and dilute phases. For monotonic $f\left(\rho\right)$ in Sec.~\ref{sec:mono}, it is always the dense and the dilute phases which are respectively ordered ($m=m_0(\rho,T)$) and disordered ($m=0$). Yet, one can have the opposite case for non-monotonic $f(\rho)$, as explained in Sec.~\ref{sec.non.mon}. For illustration purposes, considering here the case of monotonic $f(\rho)$, the condition for common tangent is given by
\begin{equation}
\begin{aligned}
\mathcal F\left[\rho_l,m_0\left(\rho_l,T\right)\right] &=\mathcal F\left(\rho_g,m=0\right)+\partial_{\rho}\mathcal F\left(\rho_g,m=0\right)\left(\rho_l-\rho_g\right) ,
\\
\partial_{\rho}\mathcal F\left[\rho_l,m_0\left(\rho_l,T\right)\right] &= \partial_{\rho}\mathcal F\left(\rho_g,m=0\right) ,
\end{aligned}
\end{equation}
which explicitly reads
\begin{equation}\label{2}
\begin{aligned}
\frac{\rho_l}{2}\left[\ln\frac{\rho_l^2\left(1-z^2\right)}{4}+z\ln\frac{1+z}{1-z}\right]-\rho_l-\beta \frac{\rho_lz^2}{2}f\left(\rho_l\right) &= \rho_g\left(\log\frac{\rho_g}{2}-1\right)+\log\frac{\rho_g}{2}\left(\rho_l-\rho_g\right) ,
\\
\log\left(\frac{\rho_l}{2}\sqrt{1-z^2}\right)-\frac{\beta z^2}{2}\left[\rho_lf'\left(\rho_l\right)-f\left(\rho_l\right)\right] &= \log\frac{\rho_g}{2} ,
\end{aligned}
\end{equation}
where we have used the parametric representation $z=m_0(\rho_l,T)/\rho_l$. Substituting the temperature relation~\eqref{tx} into~\eqref{2}, we get
\begin{equation}\label{bin}
1-\frac{\rho_g}{\rho_l} = \frac{z\text{arctanh}  z}{2}\frac{\rho_lf'\left(\rho_l\right)}{f\left(\rho_l\right)} ,
\qquad
\frac{\rho_g}{\rho_l} = \sqrt{1-z^2}e^{-\frac{z\text{arctanh}  z}{2}\left[\frac{\rho_lf'\left(\rho_l\right)}{f\left(\rho_l\right)}-1\right]}
\end{equation}
which finally gives the two binodals in an implicit parametric representation for $\rho_l\left(z\right)$, $\rho_g\left(z\right)$ and $T=T\left(z\right)$. The tricritical point is defined as the meeting point of the two binodals~\eqref{bin}, which is also the meeting point of the two spinodals~\eqref{spin} and~\eqref{inst1}. This occurs at vanishing $z=m_0/\rho_0$. Indeed, expanding either one of these expressions at small $x$ leads to~\eqref{cond0}.


\section{Linear instability in the nonequilibrium case}\label{ap.non_eq_phase}

To derive the expression of the density spinodal in~\eqref{instp}, we examine the linear stability of the dynamics~(\ref{d1}-\ref{d2}) around the homogeneous solutions $\rho\left(x,t\right)=\rho_0+\delta\rho\left(x,t\right)$ and $m\left(x,t\right)=m_0\left(\rho_0,T\right)+\delta m \left(x,t\right)$, where $m_0$ is given by~\eqref{mstar}. Using the Fourier convention for an arbitrary field $X$:
\begin{equation}
\delta X(x, t) = \sum_{n} \; \delta X_n(t) e^{ik_nx} ,
\quad k_n=2\pi n/\ell_s ,
\end{equation}
we arrive at the following mode dynamics
\begin{equation}\label{dyn5}
\frac{d}{dt}\begin{bmatrix}
{\delta\rho_n}\\
{\delta m_n}
\end{bmatrix}=
\mathbb M
\begin{bmatrix}
\delta\rho_n\\
\delta m_n
\end{bmatrix} ,
\end{equation}
where the stability matrix $\mathbb M$ reads
\begin{equation}\label{matrix}
\mathbb M= \left.\begin{bmatrix} -k^2A&-k^2B-\text{Pe}ik\\
-k^2C-\text{Pe}ik-2\alpha_2&-k^2D-2\alpha_3\end{bmatrix}\right|_{\rho_0,m_0(\rho_0,T)}
\end{equation}
with
\begin{equation}
\begin{aligned}
&A=\rho\partial_{\rho\rho}\mathcal F+m\partial_{\rho m}\mathcal F,
\quad
B=\rho\partial_{\rho m}\mathcal F+m\partial_{m m}\mathcal F ,
\quad
\alpha_2=M\partial_{\rho m}\mathcal F ,
\\
&C=m\partial_{\rho\rho}\mathcal F+\rho\partial_{\rho m}\mathcal F ,
\quad
D=m\partial_{\rho m}\mathcal F+\rho\partial_{m m}\mathcal F ,
\quad
\alpha_3=M\partial_{m m}\mathcal F ,
\end{aligned}
\end{equation}
and $M\left(\rho,m\right)=\sqrt{\rho^2+m^2}$. The eigenvalues of~\eqref{matrix} are of the form
\begin{equation}\label{det}
\Lambda_{1,2}=\frac{1}{2}\left[\text{Tr}\pm\sqrt{\text{Tr}^2-4\text{Det}}\right],
\end{equation}
where
\begin{equation}
\text{Tr}=-\left[2\alpha_3+k^2\left(A+D\right)\right] ,
\quad
\text{Tr}^2-4\text{Det}=a^2+ikb+k^2c+\mathcal O(k^3) ,
\end{equation}
with
\begin{equation}\label{abc}
a=2\alpha_3 ,
\quad
b=8\text{Pe}\,\alpha_2 ,
\quad
c=4\left[\alpha_3\left(D-A\right)+2B\alpha_2-\text{Pe}^2\right] ,
\end{equation}
yielding
\begin{equation}
\text{Re}\left[\sqrt{\text{Tr}^2-4\text{Det}}\right]=a+\frac{k^2}{a}\left(\frac{b^2}{8a^2}+\frac{c}{2}\right)+\mathcal O(k^4),
\quad
\text{Im}\left[\sqrt{\text{Tr}^2-4\text{Det}}\right]=k\frac{b}{2a}+\mathcal O(k^3) ,
\end{equation}
where we used that $\alpha_3\propto \partial_{mm}\mathcal F>0$ at the magnetized state $\left[\rho_0,m_0\left(\rho_0,T\right)\right]$. (Note that the expressions $a,b,c$ in~\eqref{abc} are not to be confused with the Model C coefficients \eqref{modelc}.) Then, the eigenvalue with the larger real part is given by
\begin{equation}\label{eign}
\Lambda_+=ik\text{Pe}\frac{\partial_{\rho m}\mathcal F}{\partial_{m m}\mathcal F}+k^2\left\{\frac{\text{Pe}^2}{2M\mathcal F_{mm}}\left[\left(\frac{\partial_{\rho m}\mathcal F}{\partial_{m m}\mathcal F}\right)^2-1\right]+\frac{\rho_0\text{Hess}\left(\mathcal F\right)}{\mathcal F_{mm}}\right\}.
\end{equation}
The condition of vanishing real part for the eigenvalue~\eqref{eign} yields the ordered spinodal~\eqref{instp}.

\section{Details on numerical solution of the hydrodynamics}\label{ap.num}
To numerically solve the hydrodynamics (\ref{d1}-\ref{d2}) we employ a finite difference scheme. 
We solve it on a segment of length $\ell_s=500$ with periodic boundary conditions. The interfacial width is $\mathcal O\left(1\right)$ and the resulting traveling wave profiles have well resolved bulk phases as seen in Figs.\ref{fig.phase1}(c), \ref{fig.phase2}(c) and \ref{fig.phase3}(c). The profiles were initiated at a homogeneous state with small sinusoidal perturbation. We have confirmed that the final traveling wave profile at steady state is robust for other choices of initial conditions. For some of these the relaxation to steady state proceeded via a traveling wave with multiple phase boundaries which coarsen over time towards a single phase-separated profile with sharp interfaces.

Generic values for space and time discretization are $dx=0.1$ and $dt=0.001$. However, at low temperatures the interfaces become increasingly sharper and finer meshes were chosen. Indeed, in the limit $\beta\to\infty$ the free energy \eqref{free} diverges. Thus, the dynamics is dominated by the equilibrium free energy derivatives entering \eqref{d2}. As noted in \ref{ap.hydro}, the equilibrium hydrodynamics becomes ill-defined for any non-convex free energy $\mathcal F$ since this free energy is missing the standard interfacial energy terms and the width of the interfacial profile will vanish. The self propulsion terms in \eqref{d2} cure this and assign a finite interface width, but this decreases with decreasing temperature and is hard to resolve numerically. Luckily, due to the same reasoning, close to vanishing temperatures the binodals of the non-equilibrium phase diagram $\text{Pe}>0$ must approach those of the equilibrium $\text{Pe}=0$ limit. As explained in \ref{ap.eq_phase}, the latter are determined analytically without the need of solving the hydrodynamics (\ref{d1}-\ref{d2}), which for the equilibrium case are ill-defined. We use these equilibrium asymptotics to extend the binodal curves of the non-equilibrium phase diagrams all the way to zero temperature (the numerically computed curves at small temperatures are found to match them smoothly).


\section*{References}

\bibliographystyle{iopart-num}
\bibliography{references}

\end{document}